%% file: main.tex
\pdfoutput=1

\documentclass[12pt,a4paper]{article}

\usepackage{ifthen} 
\newboolean{pdflatex}
\setboolean{pdflatex}{true} 

\newboolean{articletitles}
\setboolean{articletitles}{true} 

\newboolean{uprightparticles}
\setboolean{uprightparticles}{false} 

\newboolean{inbibliography}
\setboolean{inbibliography}{false} 

\def\paperauthors{LHCb collaboration} 
\def\paperasciititle{Centrality determination in heavy-ion collisions with the LHCb detector} 
\def\papertitle{Centrality determination\\ in heavy-ion collisions\\ with the LHCb detector} 
\def\paperkeywords{{High Energy Physics}, {LHCb}, {Centrality}, {Heavy Ions}, {fixed-target}} 
\def\papercopyright{\the\year\ CERN for the benefit of the LHCb collaboration} 
\def\paperlicence{CC-BY-4.0 licence}
\def\paperlicenceurl{https://creativecommons.org/licenses/by/4.0/}

\input{preamble}

\usepackage{longtable} 

\begin{document}

\renewcommand{\thefootnote}{\fnsymbol{footnote}}
\setcounter{footnote}{1}

\input{title-LHCb-PAPER}


\renewcommand{\thefootnote}{\arabic{footnote}}
\setcounter{footnote}{0}

\cleardoublepage


\pagestyle{plain} 
\setcounter{page}{1}
\pagenumbering{arabic}

%

\input{introduction}

\input{detector}

\input{glauber}

\input{data}

\input{centrality_determination}

\input{uncertainties}

\input{conclusions}

\input{acknowledgements}

\addcontentsline{toc}{section}{References}
\setboolean{inbibliography}{true}
\bibliographystyle{LHCb}
\bibliography{main,LHCb-DP}

\newpage
\input{Authorship_LHCb-DP-2021-002}

\end{document}

%% file: preamble.tex
\usepackage[top=1in, bottom=1.25in, left=1in, right=1in]{geometry}

\usepackage{microtype}
\usepackage{lineno}  
\usepackage{xspace} 
\usepackage{caption} 
\usepackage{listings}
\usepackage{xcolor}
\usepackage{subcaption}

\definecolor{codegreen}{rgb}{0,0.6,0}
\definecolor{codegray}{rgb}{0.5,0.5,0.5}
\definecolor{codepurple}{rgb}{0.58,0,0.82}
\definecolor{backcolour}{rgb}{0.95,0.95,0.92}

\lstdefinestyle{mystyle}{
    backgroundcolor=\color{backcolour},   
    commentstyle=\color{codegreen},
    keywordstyle=\color{magenta},
    numberstyle=\tiny\color{codegray},
    stringstyle=\color{codepurple},
    basicstyle=\ttfamily\footnotesize,
    breakatwhitespace=false,         
    breaklines=true,                 
    captionpos=b,                    
    keepspaces=true,                 
    numbers=left,                    
    numbersep=5pt,                  
    showspaces=false,                
    showstringspaces=false,
    showtabs=false,                  
    tabsize=2
}

\usepackage{graphicx}  
\usepackage{color}
\usepackage{colortbl}
\graphicspath{{./figs/}} 
\DeclareGraphicsExtensions{.pdf,.PDF,png,.PNG}

\usepackage{mathtools}
\usepackage{amsmath} 
\usepackage{amssymb}
\usepackage{amsfonts}
\usepackage{upgreek} 

\newcommand*\patchAmsMathEnvironmentForLineno[1]{%
\expandafter\let\csname old#1\expandafter\endcsname\csname #1\endcsname
\expandafter\let\csname oldend#1\expandafter\endcsname\csname
end#1\endcsname
 \renewenvironment{#1}%
   {\linenomath\csname old#1\endcsname}%
   {\csname oldend#1\endcsname\endlinenomath}%
}
\newcommand*\patchBothAmsMathEnvironmentsForLineno[1]{%
  \patchAmsMathEnvironmentForLineno{#1}%
  \patchAmsMathEnvironmentForLineno{#1*}%
}
\AtBeginDocument{%
\patchBothAmsMathEnvironmentsForLineno{equation}%
\patchBothAmsMathEnvironmentsForLineno{align}%
\patchBothAmsMathEnvironmentsForLineno{flalign}%
\patchBothAmsMathEnvironmentsForLineno{alignat}%
\patchBothAmsMathEnvironmentsForLineno{gather}%
\patchBothAmsMathEnvironmentsForLineno{multline}%
\patchBothAmsMathEnvironmentsForLineno{eqnarray}%
}

\usepackage{hyperxmp}

\usepackage[pdftex,
            pdfauthor={\paperauthors},
            pdftitle={\paperasciititle},
            pdfkeywords={\paperkeywords},
            pdfcopyright={Copyright (C) \papercopyright},
            pdflicenseurl={\paperlicenceurl}]{hyperref}

\usepackage[colorinlistoftodos,textsize=scriptsize]{todonotes}

\usepackage[all]{hypcap} 

\input{lhcb-symbols-def} 

\usepackage{cite} 
\usepackage{mciteplus}

%% file: lhcb-symbols-def.tex
\usepackage{xspace} 
\usepackage{upgreek}

\def\lhcb   {\mbox{LHCb}\xspace}
\def\atlas  {\mbox{ATLAS}\xspace}
\def\cms    {\mbox{CMS}\xspace}
\def\alice  {\mbox{ALICE}\xspace}

\def\velo   {VELO\xspace}

\def\spd    {SPD\xspace}

\def\ecal   {ECAL\xspace}

\def\MagUp {\mbox{\em Mag\kern -0.05em Up}\xspace}

\ifthenelse{\boolean{uprightparticles}}%
{

 \def\PDelta      {\ensuremath{\Delta}\xspace}                 
 \def\PXi         {\ensuremath{\Xi}\xspace}                 
 \def\PLambda     {\ensuremath{\Lambda}\xspace}                 
 \def\PSigma      {\ensuremath{\Sigma}\xspace}                 
 \def\POmega      {\ensuremath{\Omega}\xspace}                 
 \def\PUpsilon    {\ensuremath{\Upsilon}\xspace}

 \def\PB      {\ensuremath{\mathrm{B}}\xspace}                 
                  
 \def\PD      {\ensuremath{\mathrm{D}}\xspace}

 \def\PK      {\ensuremath{\mathrm{K}}\xspace}

 \def\Pb      {\ensuremath{\mathrm{b}}\xspace}                 
 \def\Pc      {\ensuremath{\mathrm{c}}\xspace}

 \def\Pi      {\ensuremath{\mathrm{i}}\xspace}

 \def\Ps      {\ensuremath{\mathrm{s}}\xspace}

 \def\thebaroffset{0.0em}
}
{

 \mathchardef\PDelta="7101
 \mathchardef\PXi="7104
 \mathchardef\PLambda="7103
 \mathchardef\PSigma="7106
 \mathchardef\POmega="710A
 \mathchardef\PUpsilon="7107
                  
 \def\PB      {\ensuremath{B}\xspace}                 
                  
 \def\PD      {\ensuremath{D}\xspace}

 \def\PK      {\ensuremath{K}\xspace}

 \def\Pb      {\ensuremath{b}\xspace}                 
 \def\Pc      {\ensuremath{c}\xspace}

 \def\Pi      {\ensuremath{i}\xspace}

 \def\Ps      {\ensuremath{s}\xspace}

 \def\thebaroffset{0.18em}
}
\newcommand{\offsetoverline}[2][\thebaroffset]{\kern #1\overline{\kern -#1 #2}}%

\makeatletter
\ifcase \@ptsize \relax
  \newcommand{\miniscule}{\@setfontsize\miniscule{4}{5}}
\or
  \newcommand{\miniscule}{\@setfontsize\miniscule{5}{6}}
\or
  \newcommand{\miniscule}{\@setfontsize\miniscule{5}{6}}
\fi
\makeatother

\DeclareRobustCommand{\optbar}[1]{\shortstack{{\miniscule (\rule[.5ex]{1.25em}{.18mm})}
  \\ [-.7ex] $#1$}}

\def\squark    {{\ensuremath{\Ps}}\xspace}

\def\cquark    {{\ensuremath{\Pc}}\xspace}

\def\bquark    {{\ensuremath{\Pb}}\xspace}

\def\KorKbar {\kern \thebaroffset\optbar{\kern -\thebaroffset \PK}{}\xspace}

\def\DorDbar {\kern \thebaroffset\optbar{\kern -\thebaroffset \PD}\xspace}

\def\B       {{\ensuremath{\PB}}\xspace}

\def\BorBbar {\kern \thebaroffset\optbar{\kern -\thebaroffset \PB}\xspace}

\def\Bd      {{\ensuremath{\B^0}}\xspace}

\def\BdorBdbar {\kern \thebaroffset\optbar{\kern -\thebaroffset \Bd}\xspace}

\def\Bs      {{\ensuremath{\B^0_\squark}}\xspace}

\def\BsorBsbar {\kern \thebaroffset\optbar{\kern -\thebaroffset \Bs}\xspace}

\def\Y#1S{\ensuremath{\PUpsilon{(#1S)}}\xspace}

\def\LorLbar     {\kern \thebaroffset\optbar{\kern -\thebaroffset \PLambda}\xspace}

\def\AT#1     {\ensuremath{A_{\mathrm{T}}^{#1}}\xspace}           

\def\C#1      {\ensuremath{\mathcal{C}_{#1}}\xspace}                       
\def\Cp#1     {\ensuremath{\mathcal{C}_{#1}^{'}}\xspace}                    
\def\Ceff#1   {\ensuremath{\mathcal{C}_{#1}^{\mathrm{(eff)}}}\xspace}        
\def\Cpeff#1  {\ensuremath{\mathcal{C}_{#1}^{'\mathrm{(eff)}}}\xspace}       
\def\Ope#1    {\ensuremath{\mathcal{O}_{#1}}\xspace}                       
\def\Opep#1   {\ensuremath{\mathcal{O}_{#1}^{'}}\xspace}                    

\newcommand{\nospaceunit}[1]{\ensuremath{\text{#1}}}       
\newcommand{\aunit}[1]{\ensuremath{\text{\,#1}}}       

\newcommand{\tev}{\aunit{Te\kern -0.1em V}\xspace}
\newcommand{\gev}{\aunit{Ge\kern -0.1em V}\xspace}
\newcommand{\mev}{\aunit{Me\kern -0.1em V}\xspace}
\newcommand{\kev}{\aunit{ke\kern -0.1em V}\xspace}
\newcommand{\ev}{\aunit{e\kern -0.1em V}\xspace}
\newcommand{\mevc}{\ensuremath{\aunit{Me\kern -0.1em V\!/}c}\xspace}
\newcommand{\gevc}{\ensuremath{\aunit{Ge\kern -0.1em V\!/}c}\xspace}
\newcommand{\mevcc}{\ensuremath{\aunit{Me\kern -0.1em V\!/}c^2}\xspace}
\newcommand{\gevcc}{\ensuremath{\aunit{Ge\kern -0.1em V\!/}c^2}\xspace}

\def\m    {\aunit{m}\xspace}

\def\mm   {\aunit{mm}\xspace}

\def\mum  {\ensuremath{\,\upmu\nospaceunit{m}}\xspace}

\def\fm   {\aunit{fm}\xspace}

\def\mbarn{\aunit{mb}\xspace}

\def\pb {\aunit{pb}\xspace}

\newcommand{\chisq}{\ensuremath{\chi^2}\xspace}
\newcommand{\chisqndf}{\ensuremath{\chi^2/\mathrm{ndf}}\xspace}

\def\deriv {\ensuremath{\mathrm{d}}}

\def\gsim{{~\raise.15em\hbox{$>$}\kern-.85em
          \lower.35em\hbox{$\sim$}~}\xspace}
\def\lsim{{~\raise.15em\hbox{$<$}\kern-.85em
          \lower.35em\hbox{$\sim$}~}\xspace}

\def\sqsnn {\ensuremath{\protect\sqrt{s_{\scriptscriptstyle\text{NN}}}}\xspace}
\def\pt         {\ensuremath{p_{\mathrm{T}}}\xspace}

\xspace

\def\tell1  {TELL1\xspace}
\def\ukl1   {UKL1\xspace}

\newcommand{\ie}{\mbox{\itshape i.e.}\xspace}

\def\npart	{\ensuremath{N_\mathrm{{part}}}\xspace}
\def\ncoll	{\ensuremath{N_\mathrm{{coll}}}\xspace}
\def\nanc 	{\ensuremath{N_\mathrm{{anc}}}\xspace}

\def\nout	{\ensuremath{N_\mathrm{{out}}}\xspace}
\def\b	{\ensuremath{b}\xspace}

\def\mnpart	{\ensuremath{\langle N_\mathrm{{part}}\rangle}\xspace}
\def\mncoll	{\ensuremath{\langle N_\mathrm{{coll}}\rangle}\xspace}

\def\siginel {\ensuremath{\sigma_{\mathrm{NN}}^{\mathrm{inel}}}\xspace}
\def\pb {\ensuremath{^{208}\mathrm{Pb}}\xspace}

%% file: title-LHCb-PAPER.tex

\begin{titlepage}
\pagenumbering{roman}

\vspace*{-1.5cm}
\centerline{\large EUROPEAN ORGANIZATION FOR NUCLEAR RESEARCH (CERN)}
\vspace*{1.5cm}
\noindent
\begin{tabular*}{\linewidth}{lc@{\extracolsep{\fill}}r@{\extracolsep{0pt}}}
\ifthenelse{\boolean{pdflatex}}
{\vspace*{-1.5cm}\mbox{\!\!\!\includegraphics[width=.14\textwidth]{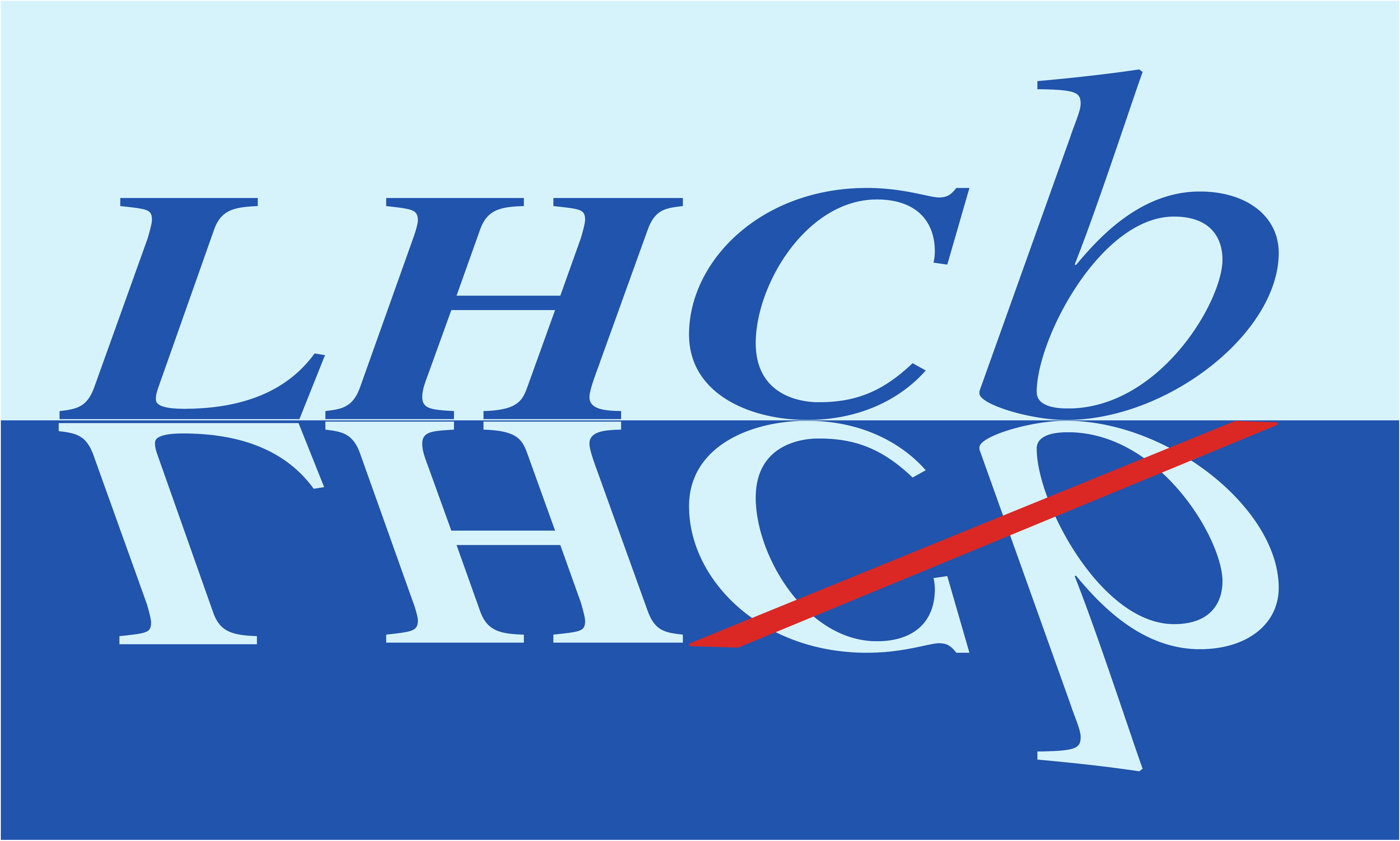}} & &}%
{\vspace*{-1.2cm}\mbox{\!\!\!\includegraphics[width=.12\textwidth]{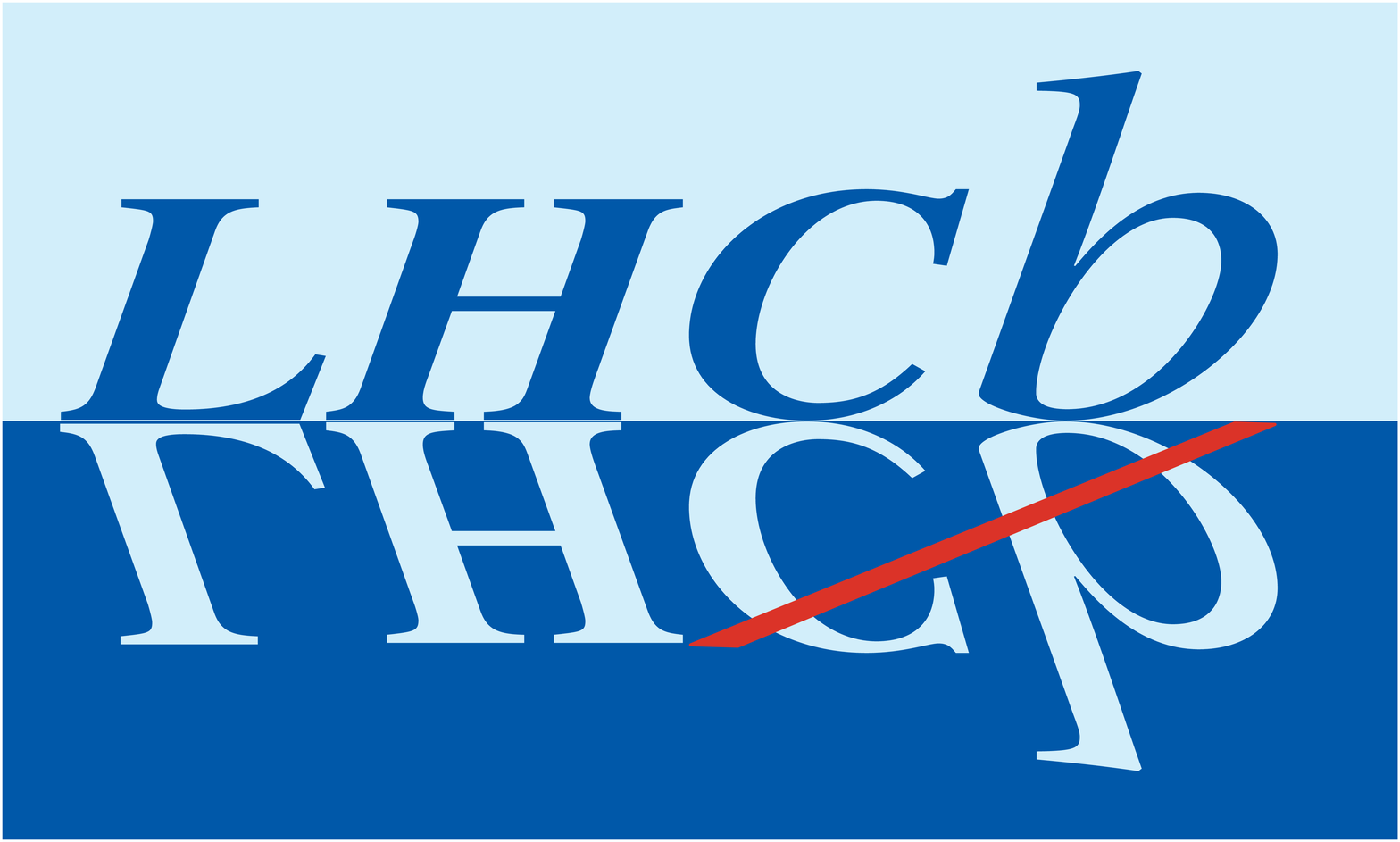}} & &}%
\\
 & & CERN-LHCb-DP-2021-002 \\  
 & & October 19, 2021 \\
 & & \\
\end{tabular*}

\vspace*{4.0cm}

{\normalfont\bfseries\boldmath\huge
\begin{center}
  \papertitle 
\end{center}
}

\vspace*{2.0cm}

\begin{center}
\paperauthors\footnote{Authors are listed at the end of this paper.}
\end{center}

\vspace{\fill}

\begin{abstract}
  \noindent The centrality of heavy-ion collisions is directly related to the medium created therein. A procedure to determine the centrality of collisions with the \lhcb detector is implemented for lead-lead collisions at $\sqsnn=5\tev$ and lead-neon fixed-target collisions at $\sqsnn=69\gev$. The energy deposits in the electromagnetic calorimeter are used to determine and define the centrality classes. The correspondence between the number of participants and the centrality for the lead-lead collisions is in good agreement with the correspondence found in other experiments, and the centrality measurements for the lead-neon collisions presented here are the first performed in fixed-target collisions at the LHC.
  
\end{abstract}

\vspace*{2.0cm}

\begin{center}
 Submitted to
  J. Instr. 
\end{center}

\vspace{\fill}

{\footnotesize 
\centerline{\copyright~\papercopyright. \href{\paperlicenceurl}{\paperlicence}.}}
\vspace*{2mm}

\end{titlepage}


\newpage
\setcounter{page}{2}
\mbox{~}
%
%
%
%

\cleardoublepage

%% file: introduction.tex
\section{Introduction}
\label{sec:Introduction}

\begin{figure}[bp]
  \centering
  \includegraphics[width=\textwidth]{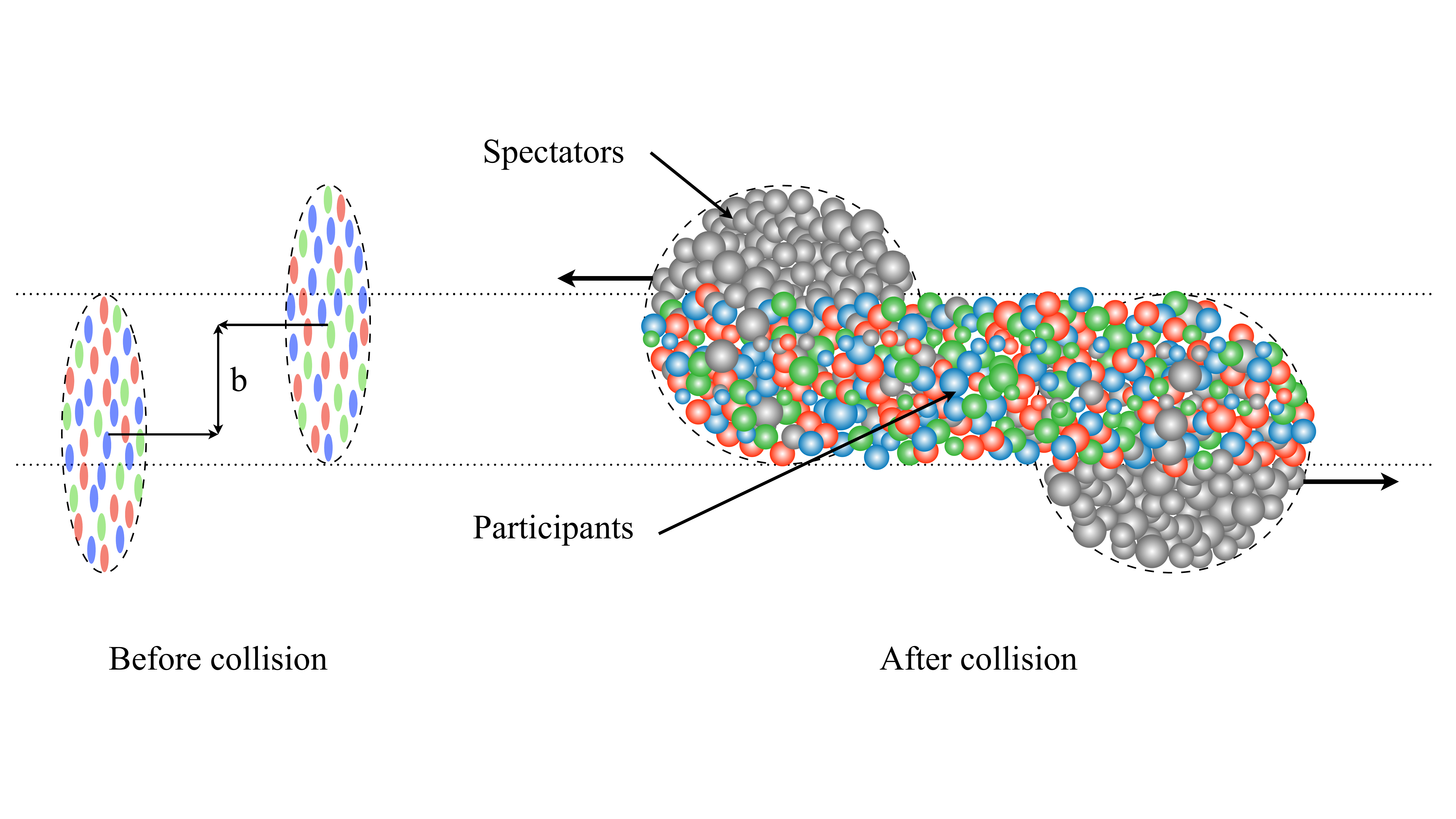}
  \caption{A schematic view of a heavy-ion collision. The impact parameter $b$ is shown as well as the spectator nucleons and the participant nucleons.}
  \label{fig:heavy-ion}
  \end{figure}

In the context of heavy-ion collisions, centrality is a quantity of relevance since it is directly related to the medium formed by the colliding nuclei, and measures the overlap region between the two nuclei in a collision. The centrality of a collision is characterised by the impact parameter ($b$) between the two nuclei, \ie the distance between their centres in the plane transverse to the beam axis. The impact parameter defines the overlap region of the nuclei and thus determines also the size and shape of the resulting medium. A schematic view of a heavy-ion collision is shown in Fig.~\ref{fig:heavy-ion}.

The geometry of the collision is related to the number of nucleons that participate in it and the number of nucleon-nucleon collisions. These quantities are not directly accessible and hence need to be derived from the data recorded during the collisions by making use of  other quantities that scale approximately with the number of participating nucleons, such as the outgoing particle multiplicity. For this purpose, a Glauber model is often used~\cite{Miller2007ri}.
 
This paper presents the centrality determination with the LHCb detector for lead-lead (PbPb) collisions at a centre-of-mass energy $\sqsnn = 5\tev$, which is in agreement with results obtained by the \alice~\cite{PhysRevLett.116.222302, ALICE-PUBLIC-2015-008}, \atlas~\cite{ATLAS2019108} and \cms~\cite{PhysRevLett.127.102002} collaborations, and the first centrality determination in fixed-target mode at the LHC, for lead-neon (PbNe) collisions at $\sqsnn = 69\gev$. After introducing the Glauber model (section~\ref{sec.glauber}) and the assumptions it relies on, the datasets (section~\ref{sec.data}), the centrality determination procedure with its results (section~\ref{sec.centrality}) and the study of systematic uncertainties (section~\ref{sec.uncertainties}), for PbPb and for PbNe collisions are described.

%% file: detector.tex
\section{The \lhcb detector}
\label{sec:Detector}

The \lhcb detector~\cite{LHCb-DP-2008-001,LHCb-DP-2014-002} is a single-arm forward spectrometer covering the \mbox{pseudorapidity} range $2<\eta <5$, designed for the study of particles containing \bquark or \cquark quarks. The detector includes a high-precision tracking system consisting of a silicon-strip vertex detector (\velo) surrounding the beam interaction region with two pile-up (PU) stations upstream from the interaction point, a large-area silicon-strip detector located upstream of a dipole magnet with a bending power of about $4{\mathrm{\,Tm}}$, and three stations of silicon-strip detectors and straw drift tubes placed downstream of the magnet. The tracking system provides a measurement of the momentum of charged particles with a relative uncertainty that varies from 0.5\% at low momentum to 1.0\% at 200\gevc. The minimum distance of a track to a primary vertex (PV), is measured with a resolution of $(15+29/\pt)\mum$, where \pt is the component of the momentum transverse to the beam, in\,\gevc. Different types of charged hadrons are distinguished using information from two ring-imaging Cherenkov detectors. Photons, electrons and hadrons are identified by a calorimeter system consisting of scintillating-pad (\spd) and pre-shower detectors, an electromagnetic (\ecal) and a hadronic calorimeter. Muons are identified by a system composed of alternating layers of iron and multiwire proportional chambers. The online event selection is performed by a trigger, which consists of a hardware stage, based on information from the calorimeter and muon systems, followed by a software stage, which applies a full event reconstruction.

A gas-injection system for beam-gas interactions (SMOG)~\cite{SMOG} is installed in the LHCb detector,
which gives the unique possibility of injecting a low pressure noble gas and collecting fixed-target collisions (proton-nucleus or nucleus-nucleus). 

%% file: glauber.tex
\section{Glauber model}
\label{sec.glauber}

\begin{figure}[tbp]
  \centering
  \includegraphics[width=0.49\textwidth]{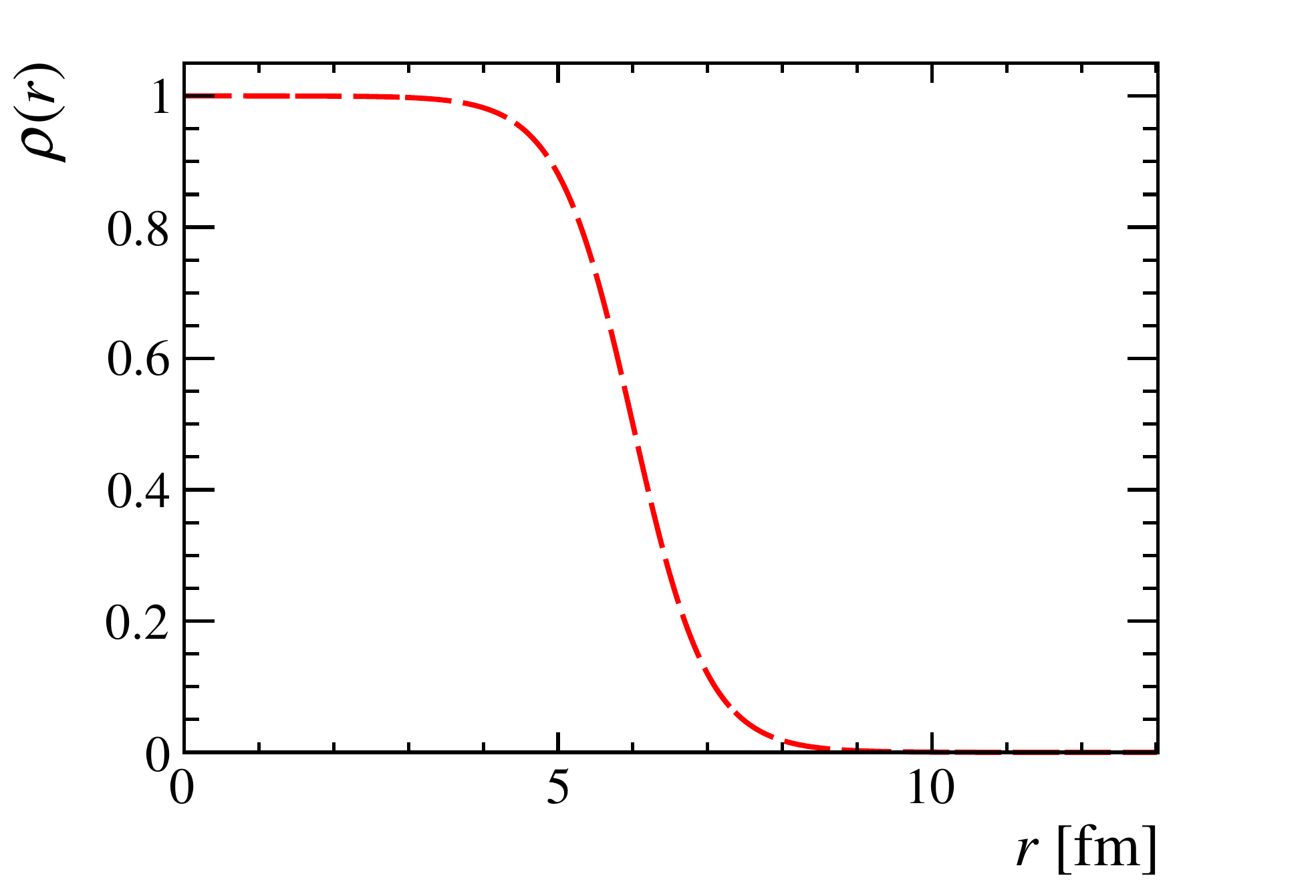}
  \caption{The 2pF density distribution $\rho$ as a function of the radius $r$. Here $w$ has been set to 0, $R=6\fm$ and $a=0.5\fm$.}
  \label{fig:density}
  \end{figure}

The centrality of a nucleus-nucleus collision is related to the overlap region between the nuclei where the nucleons are colliding.
In practice, the particles produced by the collisions are not originating purely from hadronic interactions between the nuclei, but also from electromagnetic processes. Therefore, a model is needed to isolate the hadronic part and subsequently define the centrality classes.
The most common approach in heavy-ion physics to model the collisions of two nuclei is to consider the transverse shapes of the nuclei. This shape, \ie the nuclear density, is described by a two-parameter Fermi distribution (2pF), also known as Woods--Saxon distribution~\cite{Miller2007ri}, for each nuclear species considered, defined as
\begin{equation}
\rho(r) \,\mathrm{d}r=\rho_0 \frac{1+w\,\frac{r^2}{R^2}}{1+\exp(\frac{r-R}{a})}\,\mathrm{d}r,
\label{eq.2pf}
\end{equation}
where $r$ stands for the radial distance from the centre of the nucleus.
The constant $\rho_0$ corresponds to the density at the centre of the nucleus, and $R$ to the nuclear radius, which is approximately the radial extension of the bulk of the nucleus. The diffusivity $a$ describes how abruptly the density falls at the edge of the nucleus. The last parameter, $w$, is used to describe nuclei whose maximum density is reached at a radius $r>0$. The values of these parameters are taken from other experiments, typically involving lepton-nucleus collisions and other types of nuclear spectroscopy~\cite{nucleardata,ANGELI201369}. The 2pF distribution can be seen in Fig.~\ref{fig:density} with parameters $R=6\fm$, $a=0.5\fm$ and $w=0$ for illustration.

The Glauber model is generally approached in two ways, the optical Glauber model and the Monte Carlo (MC) Glauber model. The two colliding nuclei are labelled $A$ and $B$. In the optical model, it is considered that nucleons from projectile $A$ see the target $B$ as a continuous distribution, which is described by an analytical function, and vice versa. This is also called the optical limit approximation. Subsequently, the overlap area, the number of participating nucleons (\npart) and the number of binary nucleon-nucleon collisions (\ncoll) can be obtained analytically.

On the other hand, in the MC Glauber model, the calculation is performed through a MC method where nucleons from each nucleus $A$ and $B$ are generated as hard spheres\footnote{There is a variation of the model that can also take into account the sub-nucleonic dynamics called Glauber--Gribov~\cite{Loizides:2014vua}.} and are placed around the respective centres of the nuclei following the 2pF distributions. Then a random impact parameter $b$ is sampled from the distribution $\deriv \sigma / \deriv b = 2\pi b$. Finally the nuclei are made to collide, with the following assumptions:

\begin{itemize}
\item nucleus-nucleus collisions are considered to be a superposition of several independent nucleon-nucleon collisions;
\item nucleons are treated as hard spheres moving in straight lines all along the process, even if they have undergone a collision;
\item nucleons have a geometrical transverse cross-section (\siginel) and two nucleons collide if the transverse distance between their centres is $d < \sqrt{\siginel/ \pi}$.
\end{itemize}

The average values of the number of participating nucleons \mnpart, of binary collisions \mncoll and other quantities, are obtained with a Monte Carlo simulation. The distributions of \npart, \ncoll and other quantities of interest can then be obtained for any centrality interval.

The Glauber model has two relevant external inputs, the nucleon-nucleon inelastic cross-section \siginel and the spatial distribution given by the 2pF distribution with its parameters $R$, $a$ and $w$. The cross-section is obtained from a phenomenological parametrisation tuned on data, using measurements from a broad range of energies from $\sim 20\gev$ to $\sim 60\tev$, given by $\siginel (s)=A + B \,\ln^2(s)$, with $A=25.0 \pm0.9$ and $B= 0.146\pm 0.004$ \cite{PhysRevC.97.054910}. The optical approach describes fairly well the collision process but does not completely capture the physics of the total cross-section and leads to distortions in the estimation of \npart and \ncoll compared to the estimation made with the MC approach~\cite{Miller2007ri}. Therefore, in the following, the MC Glauber model is used.

%% file: data.tex
\section{Data}
\label{sec.data}

For the centrality determination a minimum bias (MB) data sample is needed, that is, data that have the minimum possible number of selections applied, to not bias the sample.

\subsection{PbPb collisions}

\begin{figure}[tbp]
\includegraphics[width=0.5\linewidth]{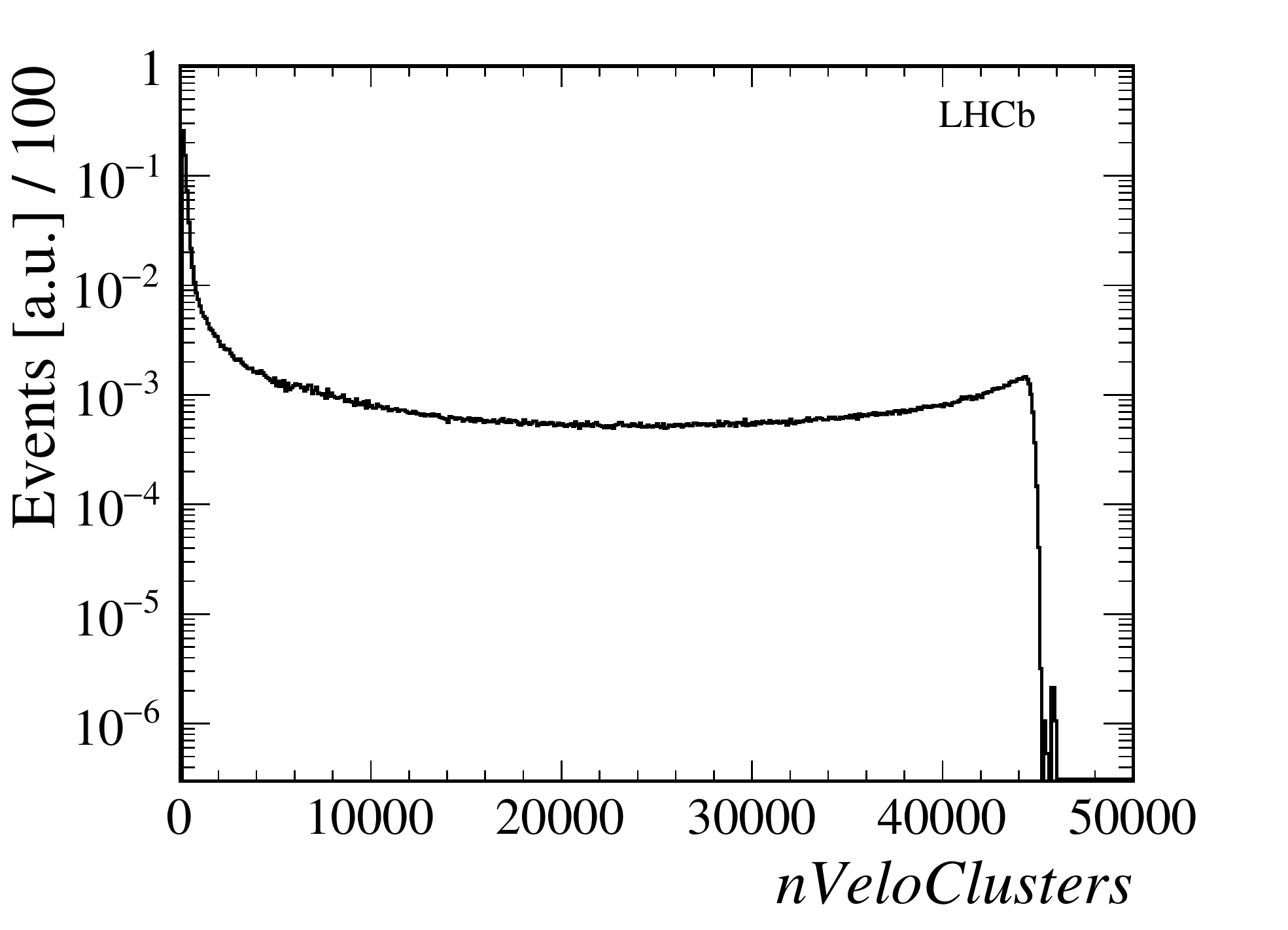}
\includegraphics[width=0.5\linewidth]{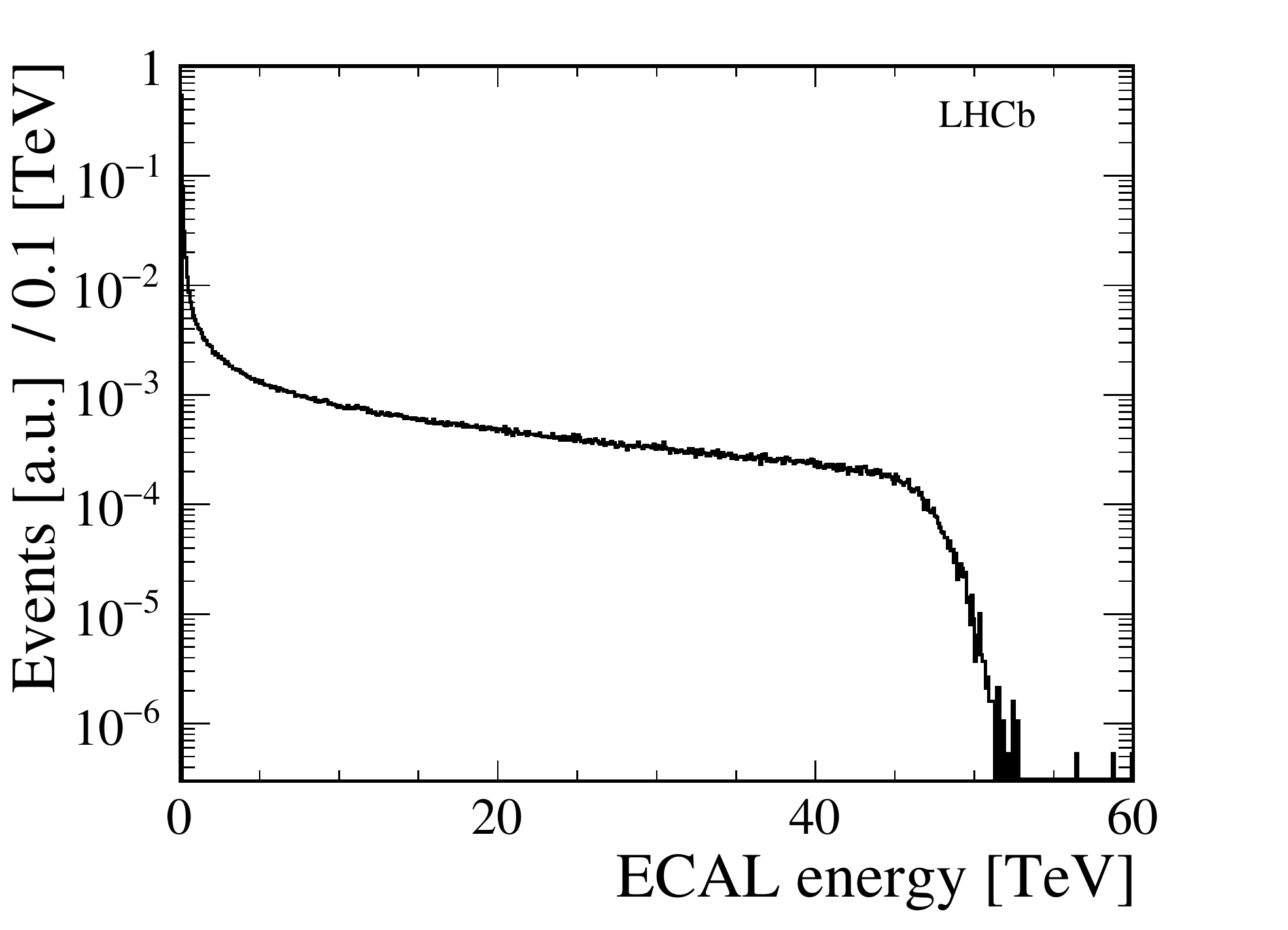}
\caption{(left) Number of \velo clusters  
and (right) energy deposited in the \ecal in PbPb collisions. The distribution of the \velo clusters exhibits a peak structure with a sharp fall at 45\,000 clusters. This is related to the total number of readout channels in the VELO, leading to saturation for high occupancy events.}
\label{fig:veloandecal}
\end{figure}

For PbPb collisions, the sample used for this analysis corresponds to the data recorded in a special run of the 2018 PbPb data-taking period, at a centre-of-mass energy per nucleon of $\sqsnn = 5\tev$. Throughout the data-taking period, PbPb collisions were recorded in parallel with fixed-target PbNe collisions, between Pb beams and atoms of Ne injected in the \velo with the SMOG system. 
However, for this particular MB data sample, no gas was injected. The resulting distribution of the number of \velo clusters (\emph{nVeloClusters}), which are clustered energy deposits in the \velo stations, and of the energy deposited in the \ecal for this MB sample can be seen in Fig.~\ref{fig:veloandecal}.

\subsection{PbNe collisions}

The PbNe data sample corresponds to a MB sample from the whole data-taking period when gas was injected. The PbNe collisions occur at a centre-of-mass energy per nucleon of $\sqsnn = 69\gev$ where the Ne atoms act as a fixed target. In order to avoid contamination from PbPb collisions that were recorded simultaneously, only events for those bunch crossings where a filled bunch from the incoming Pb beam crosses an empty bunch of the outgoing Pb beam are selected. There is some residual contamination in the data sample and extra selections are applied to increase the fraction of PbNe collisions.

The different topology of the PbPb and PbNe events allows to disentangle these two types of events by setting an upper limit on the number of clusters in the PU stations, which are located upstream from the nominal interaction point. Since the PbNe collisions are all boosted downstream, \ie towards the detector, naturally a low number of clusters in the PU stations is expected. 
On the other hand, since PbPb collisions are symmetric, a larger number of clusters in the PU stations is observed for these collisions.

Another source of contamination are the beam-gas collisions that take place far upstream. Since the injected gas can travel up to 20\m in either direction from the nominal interaction point, the incoming Pb beam can undergo interactions with the gas before arriving into the \velo tank. These events can produce forward particles, hitting the PU stations and depositing energy in the detector. 

\begin{figure}
\includegraphics[width=0.5\linewidth]{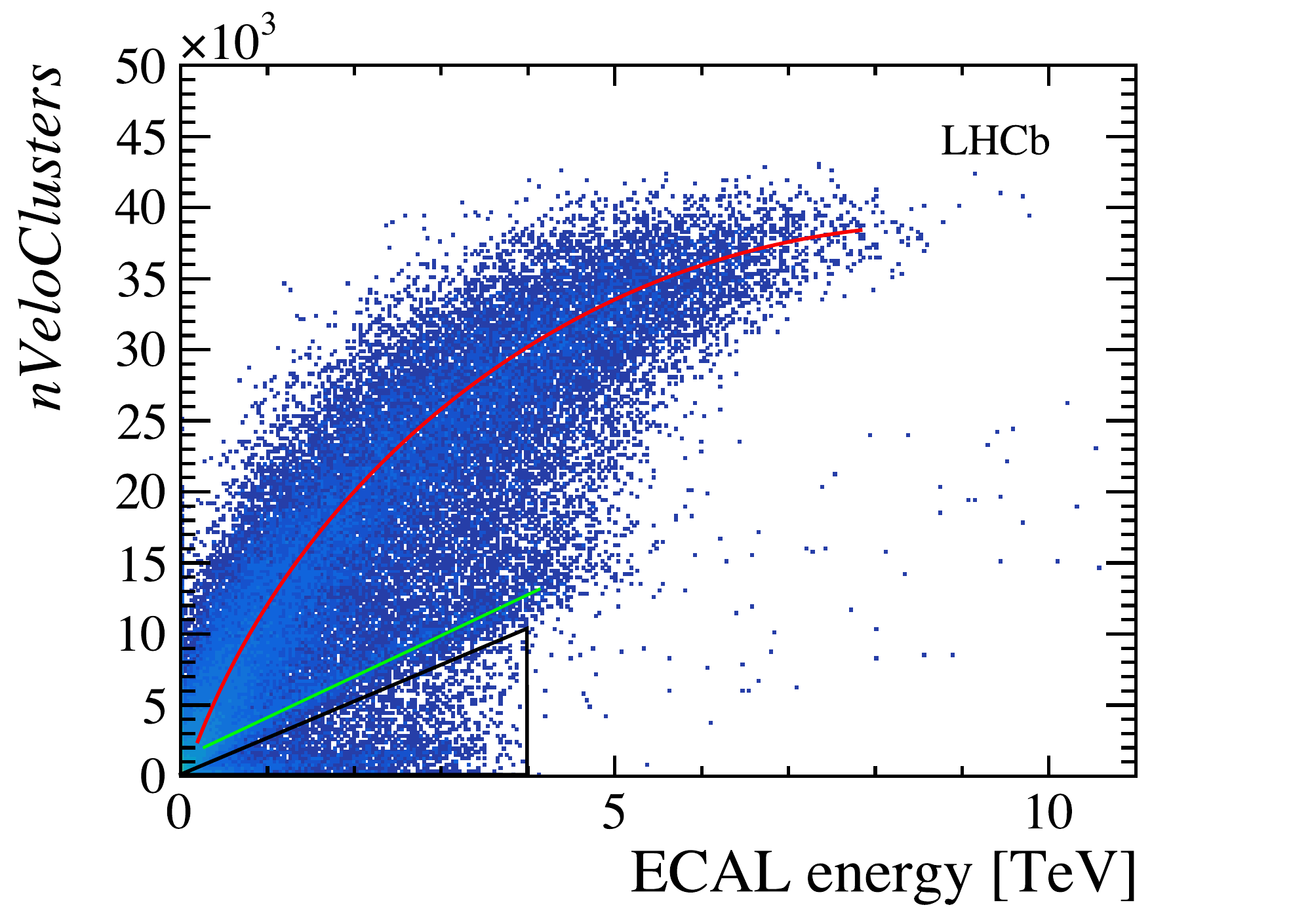}
\includegraphics[width=0.5\linewidth]{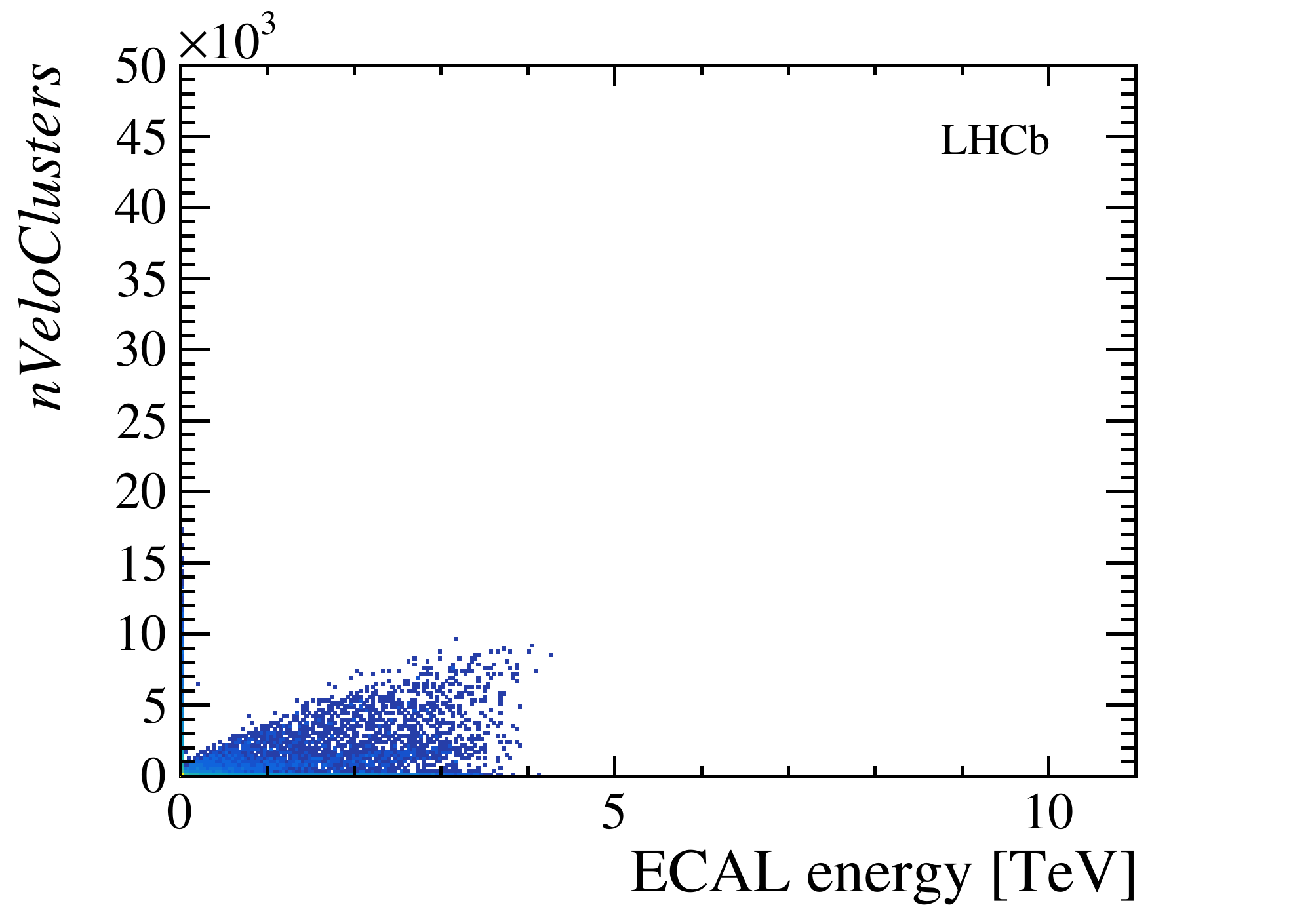}
\caption{Number of \velo clusters as a function of \ecal energy for PbNe events without any requirement (left) and without any cluster in the PU stations (right). The red line indicates the population which corresponds to the very upstream events, the green line indicates the population which corresponds to ghost PbPb collisions and the black lines enclose the PbNe collisions of interest which present no clusters in the PU stations.}
\label{fig:nvc_ecal_planes}
\end{figure}

To ensure a high enough purity of PbNe collisions in the sample, events with clusters in the PU stations are rejected. The effect of this requirement can be seen in Fig.~\ref{fig:nvc_ecal_planes}. On the left, three different populations can be seen, the high-slope population which corresponds to the very upstream events (indicated by the red line), the middle-slope population which corresponds to ghost PbPb collisions (indicated by the green line),\footnote{Ghost PbPb collisions occur when Pb ions in the downstream beam escape their nominal bunches within the beam and travel with empty outgoing bunches, leading to collisions when there should be none.} and finally the continuum which corresponds to the PbNe collisions of interest which present no clusters in the PU stations (enclosed by the black lines).

\begin{figure}
\includegraphics[width=0.5\linewidth]{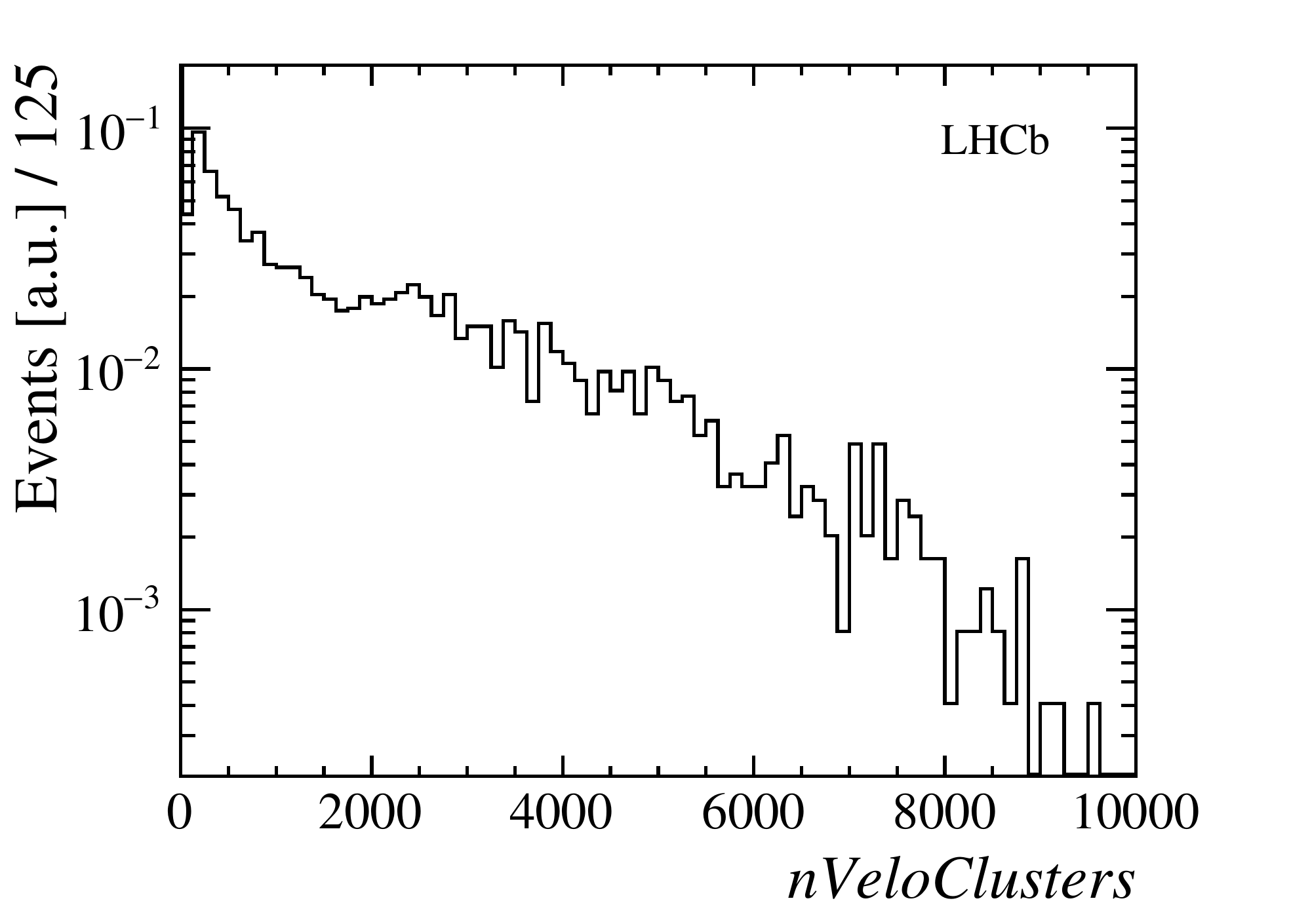}
\includegraphics[width=0.5\linewidth]{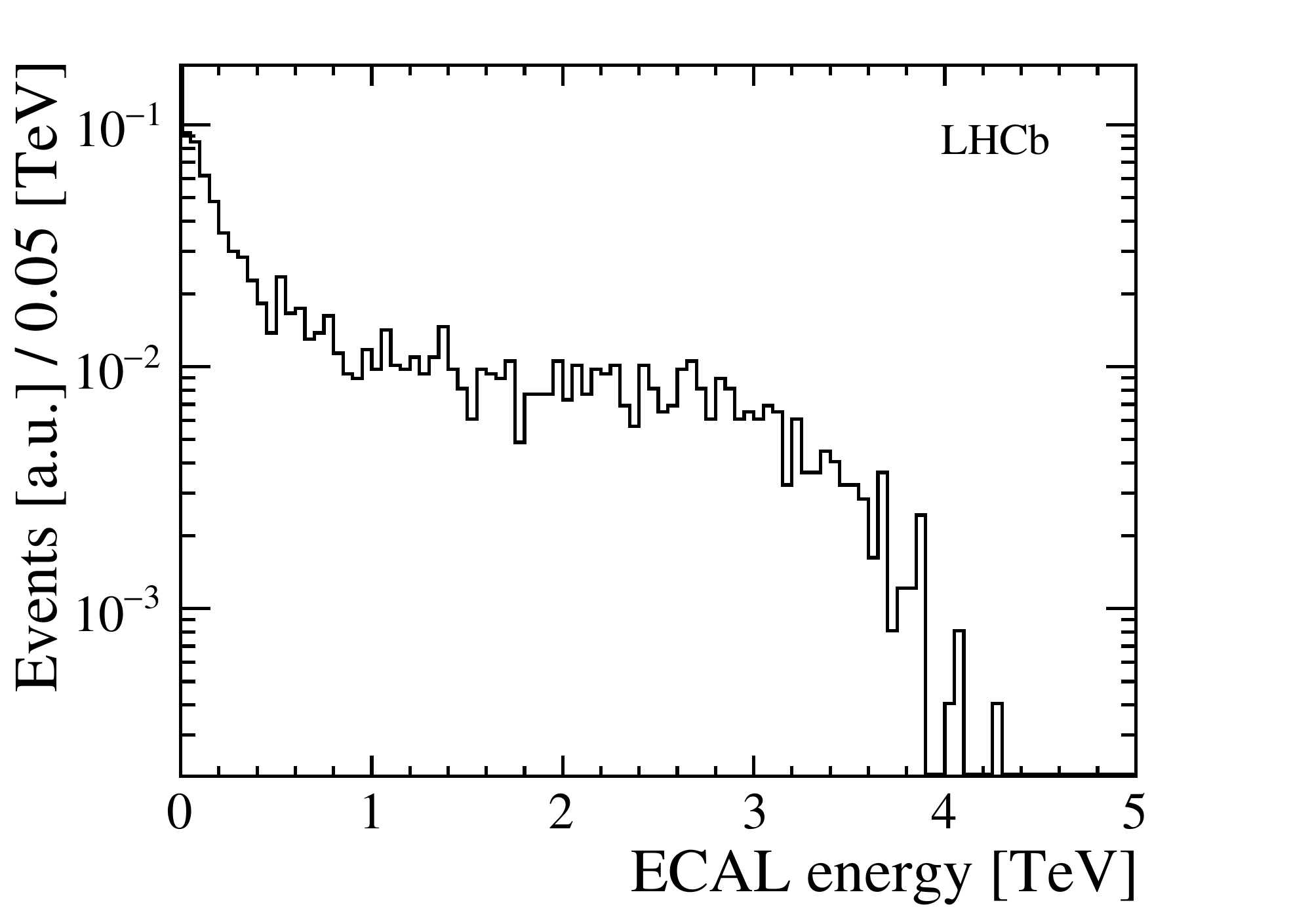}
\caption{(left) Number of \velo clusters and (right) energy deposited in the \ecal from PbNe collisions.}
\label{fig:veloandecal_pbne}
\end{figure}

Only the central-region PbNe events are used in what follows, \ie events whose primary vertex is located in the range $z_{\rm PV}\in[-200,200]\mm$. The number of \velo clusters and distribution of \ecal energy are shown in Fig.~\ref{fig:veloandecal_pbne}.

%% file: centrality_determination.tex
\section{Centrality determination}
\label{sec.centrality}

In theory, any observable that scales monotonically with impact parameter could be used for classification according to centrality. In practice, the reach in centrality possible with the tracking detectors of LHCb is limited by the performance of these detectors at high track multiplicities. In the case of PbPb collisions this means that the \velo information cannot be used for this purpose, since its sensors saturate under these conditions as seen in the rightmost part of the plot on the left of Fig.~\ref{fig:veloandecal}. In contrast, the \ecal has the advantage of not saturating even in the most central collisions, as can be seen in the right plot of Fig.~\ref{fig:veloandecal}. In the case of PbNe collisions, the \velo does not saturate but still cannot be used for the centrality determination, since the relevant events take place all along the length of the \velo. This means that the measured \velo multiplicity depends on the position of the collisions along the beam axis. For this reason the energy deposited in the \ecal is used for the multiplicity determination of both PbPb and PbNe collisions.

Centrality classes are defined as quantiles of the inelastic PbPb or PbNe cross-section. The data contain contributions to the deposited energy in the \ecal from both hadronic and electromagnetic origins (the latter originating from peripheral collisions where the electromagnetic interactions dominate and ultra-peripheral collisions (UPC)). Thus, the energy spectrum cannot be used directly to define the desired quantiles for centrality. To estimate the hadronic component, a MC Glauber model~\cite{PhysRevC.97.054910} is used to simulate the colliding nuclei, and from the resulting quantities such as \b, \ncoll or \npart, the expected observable can be constructed, which is in this case the energy deposited in the \ecal. The parameters of the model are then tuned to fit the \ecal energy distribution from the data. Finally, the centrality quantiles are defined from the simulated distribution that corresponds only to the hadronic part of the interaction. Geometric quantities from the Glauber MC can then be mapped to the data for each centrality class.

\subsection{Methodology}

In this section  the simulation of the events is described first, then the generation of the simulated \ecal energy distribution and the steps to fit it to the data are explained. Once the fit has been performed, the simulated distribution is split into centrality classes based on the fraction of the total hadronic distribution integral, and the geometric variables of each class are mapped to the measured events falling in the same class.

\subsubsection*{Events simulation}
\label{sec:sim_evt_pbpb}

The first step is to simulate the collisions using the \textsc{TGlauberMC} software from Ref.~\cite{PhysRevC.97.054910}.\footnote{For this work, in the Glauber software the Pb nucleus was specified as \texttt{Pbpnrw}, which considers slightly different distributions for protons and neutrons in the nucleus, and a reweighting of the nucleons positions to make the centre-of-mass coincide with the nominal position of the nucleus.} The parameters for the 2pF density function are listed in Table~\ref{tab:2pf_params}. In all cases, the parameter $w$ is assumed to be equal to 0.

\begin{table}[]
\centering
\caption{Parameters for the 2pF density function.}
\label{tab:2pf_params}
\begin{tabular}{lll}
\hline
                        & $R$ [\fm] & $a$ [\fm] \\ \hline
\pb                     &          &          \\
\multicolumn{1}{c}{$p$} & 6.68     & 0.45     \\
\multicolumn{1}{c}{$n$} & 6.69     & 0.56     \\
$^{20}\textrm{Ne}$      & 3.01~\cite{nucleardata,ANGELI201369}     & 0.54~\cite{PhysRevC.97.054607, doi:10.1142/3530, Seif_2015}     \\ \hline
\end{tabular}
\end{table}

One million PbPb collisions are simulated using the corresponding nucleon-nucleon cross-section $\siginel = 67.6 \pm 0.6 \mbarn$ for a centre-of-mass energy of $\sqsnn = 5 \tev$, and one million PbNe collisions were simulated using the corresponding nucleon-nucleon cross-section $\siginel = 35.4 \pm 0.9 \mbarn$ for a centre-of-mass energy of $\sqsnn = 69 \gev$.

\begin{figure}[tbp]
    \centering
    \includegraphics[width = 0.32\textwidth]{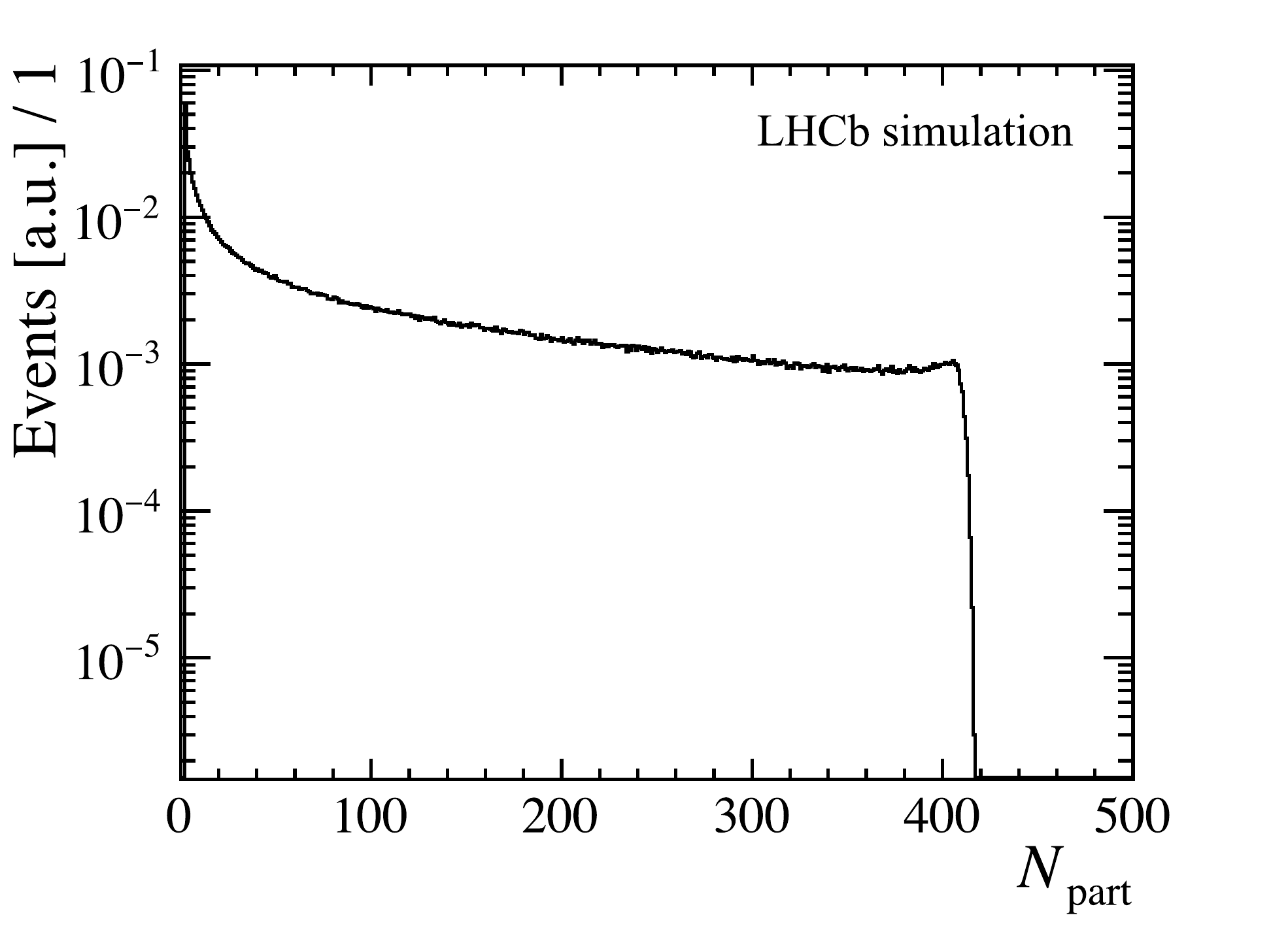}
    \includegraphics[width = 0.32\textwidth]{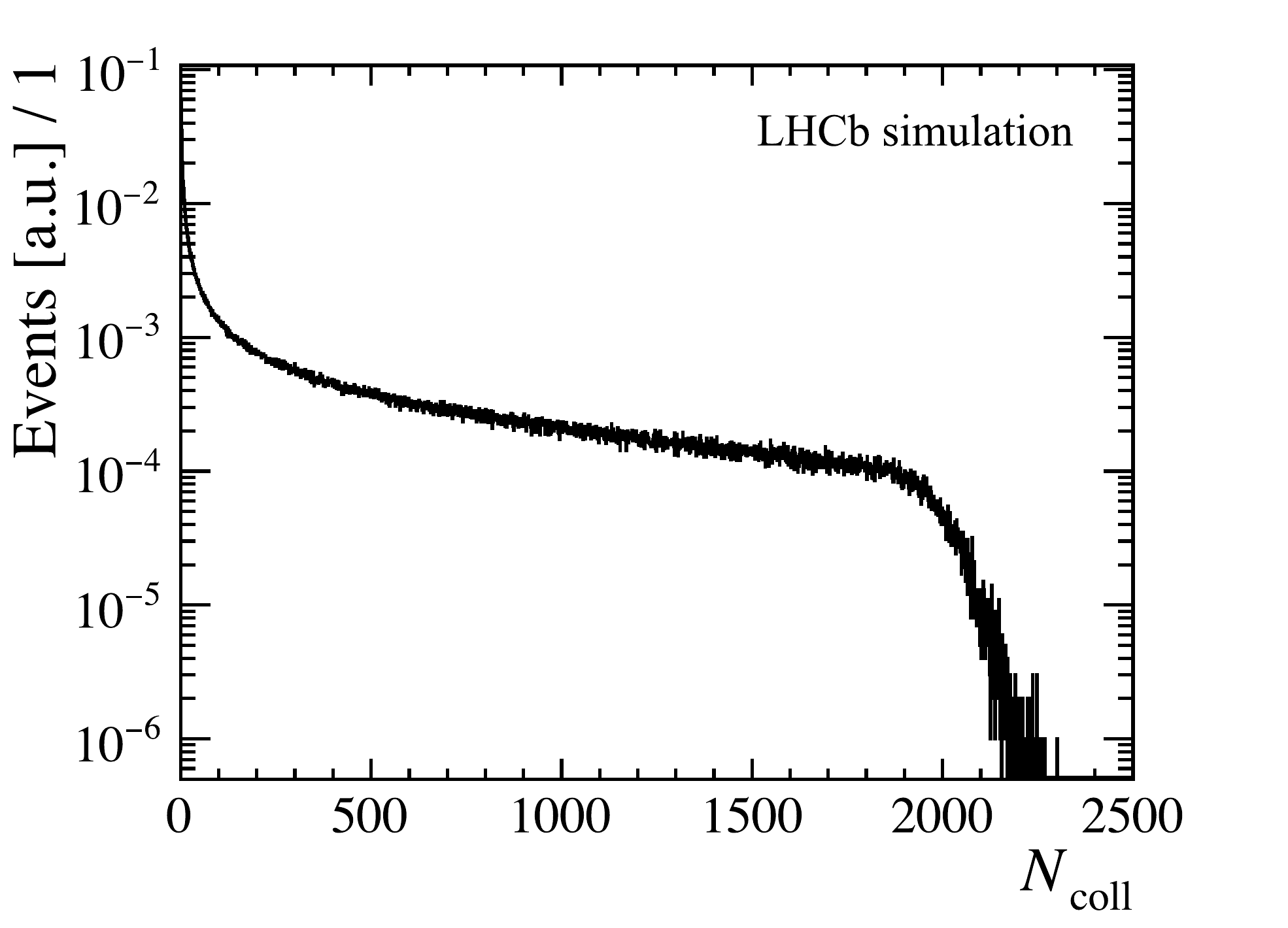}
    \includegraphics[width = 0.32\textwidth]{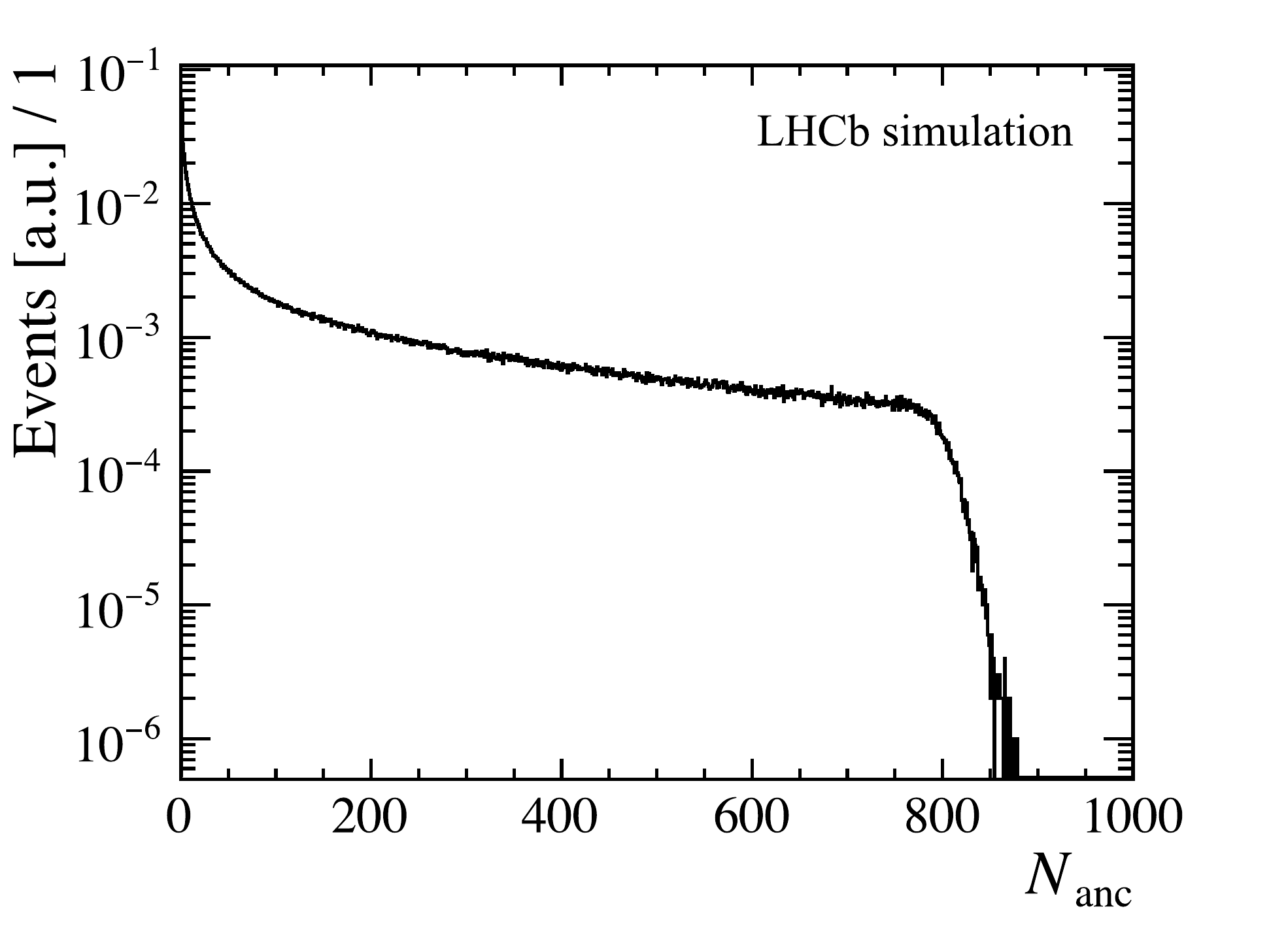}
    \caption{Distribution of (left) \npart, (middle) \ncoll and (right) \nanc from the MC Glauber model for PbPb. For this \nanc distribution, a value of $f=0.751$ was used.}
    \label{fig:glaubergeo}
  \end{figure}

The \npart and \ncoll values of every simulated collision
is then computed. With these numbers,  the number of ancestors \nanc is defined as

\begin{equation}
\nanc = f \times \npart + (1-f)\times\ncoll,
\label{eq:nanc}
\end{equation}
which effectively scales with the number of sources of particle production, with a relative weight for the participating nucleons and the number of collisions. This is motivated by the fact that the particle multiplicity is expected to scale with \npart when soft processes dominate and to scale with \ncoll when hard processes dominate~\cite{PhysRevLett.58.303,PhysRevD.38.3394,PhysRevD.39.179,PhysRevD.43.104,PhysRevC.83.014915}. Below a centre-of-mass energy of 100\gev soft processes are expected to dominate. The parameter $f$ determines the fraction of soft processes that contribute to the particle production
and has to be determined with a fit. In Fig.~\ref{fig:glaubergeo}, the distributions of \npart, \ncoll and \nanc in simulation are displayed with, as an example, $f=0.751$ for the PbPb case.

\begin{figure}[tbp]
    \centering
      \includegraphics[width=0.49\textwidth]{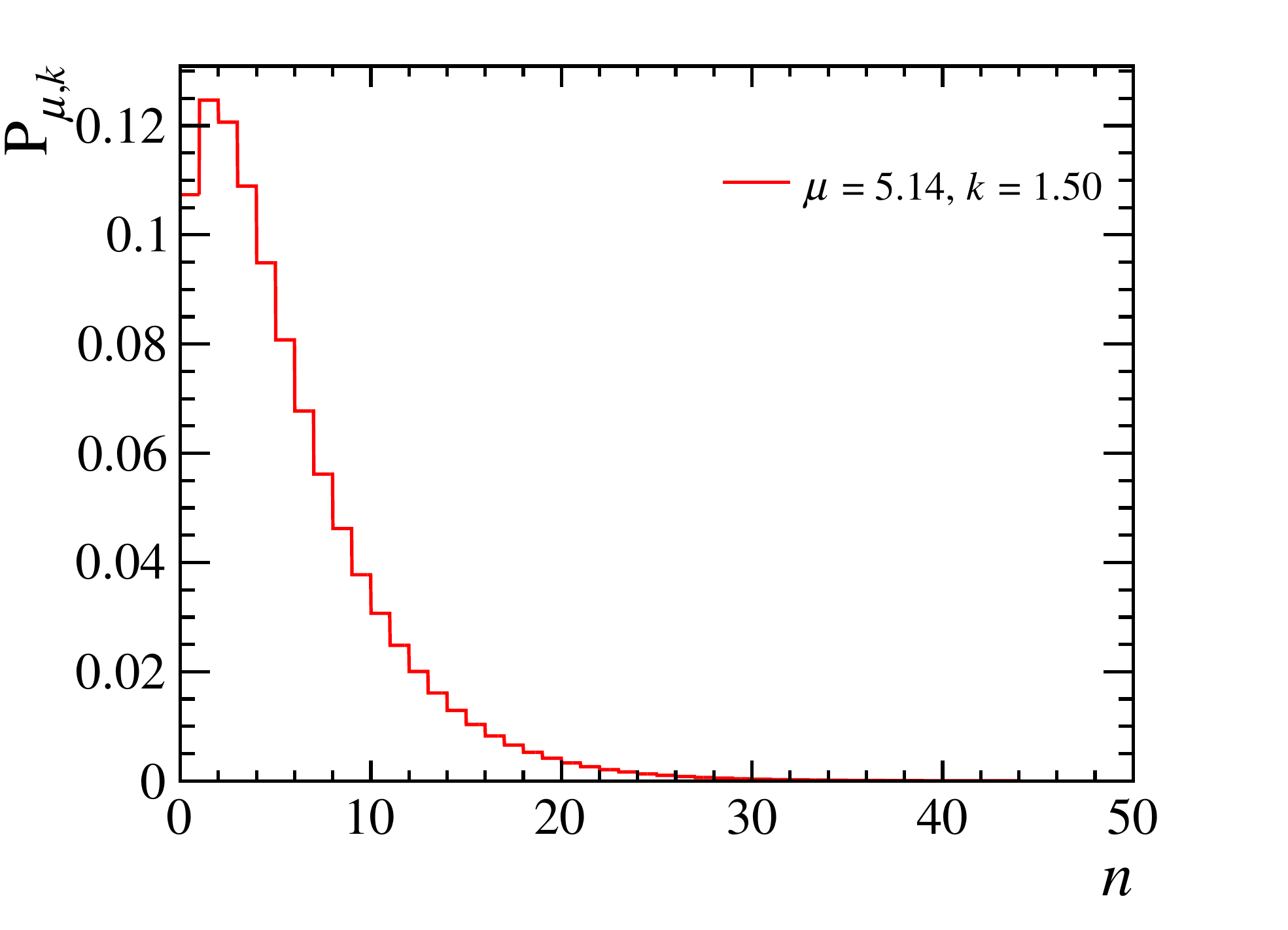}
      \includegraphics[width=0.49\textwidth]{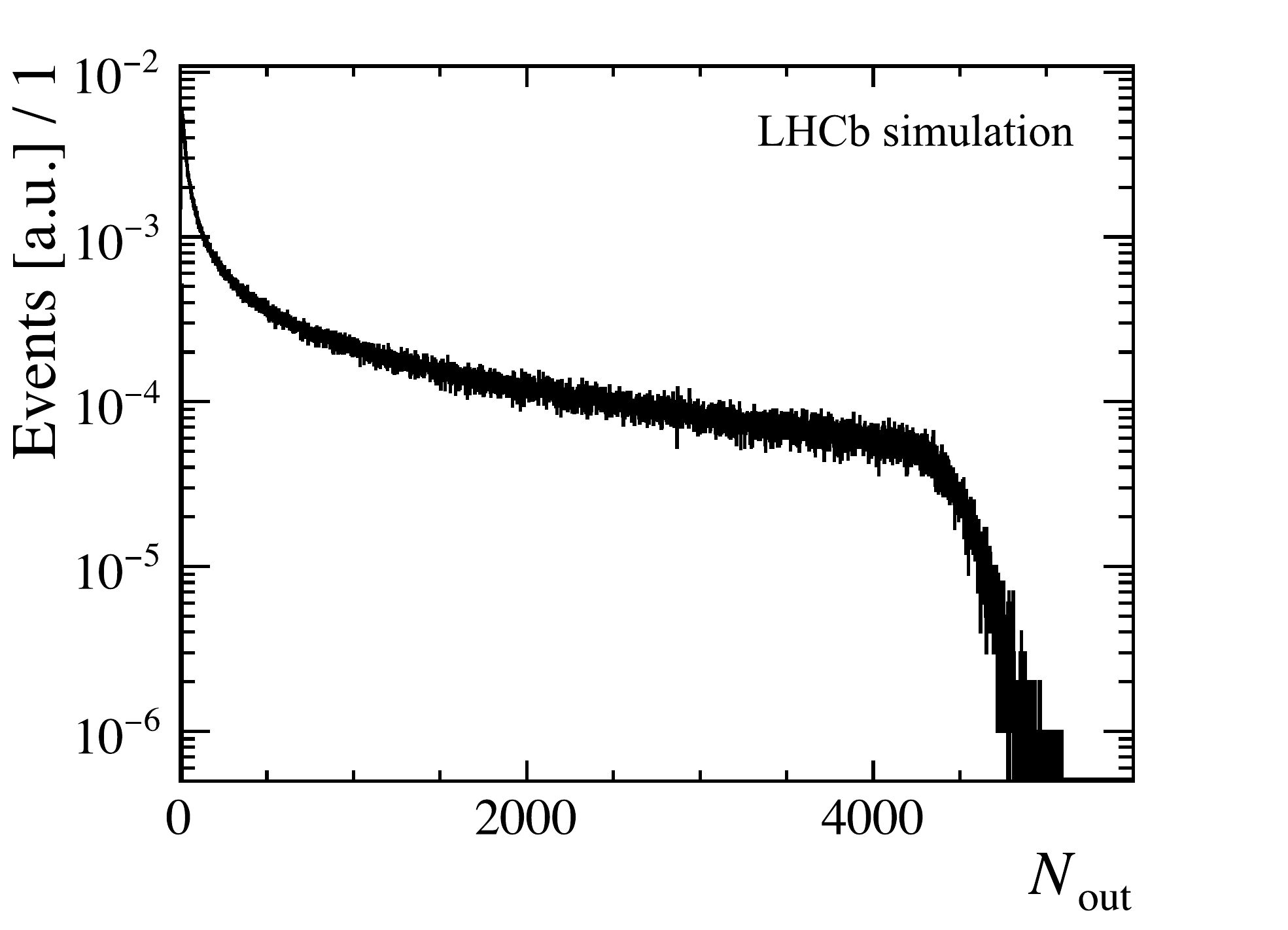}
    \caption{(left) Negative Binomial Distribution and (right) distribution of the number of outgoing particles from the MC Glauber model in PbPb collisions at $\sqsnn = 5\tev$.}
    \label{fig:nbd}
  \end{figure}

 To get the distribution of particles originating from the collision, \nanc is convoluted with a negative binomial distribution (NBD) which has been extensively used to model particle production and has been shown to be a reasonable approach at diverse energy and rapidity regimes~\cite{Cugnon_1987, Ghosh:2012xh, Aamodt:2010ft, Aamodt:2010pp, Aaij:2014pza}. The NBD is given in its discrete form by
  \begin{equation}
    P_{p,k}(n) = \frac{(n+k-1)!}{n!(k-1)!}p^k(1-p)^n,
    \label{eq:nbd}
  \end{equation}
with $p=\left(\frac{\mu}{k}+1\right)^{-1}$, where $\mu$ and $k$ are parameters related to the mean and spread of the NBD respectively, and $n$ is the number of particles that are produced and deposited energy in the \ecal. Since there are \nanc particle sources for each nucleus-nucleus collision, each producing particles following an NBD, the NBD is sampled \nanc times to get the particle multiplicity distribution. Figure~\ref{fig:nbd} illustrates the NBD function and the result after sampling it \nanc times for each event to obtain the distribution of the number of outgoing particles (\nout) which deposit energy in the \ecal.

\begin{figure}[tbp]
    \centering
      \includegraphics[width=0.49\textwidth]{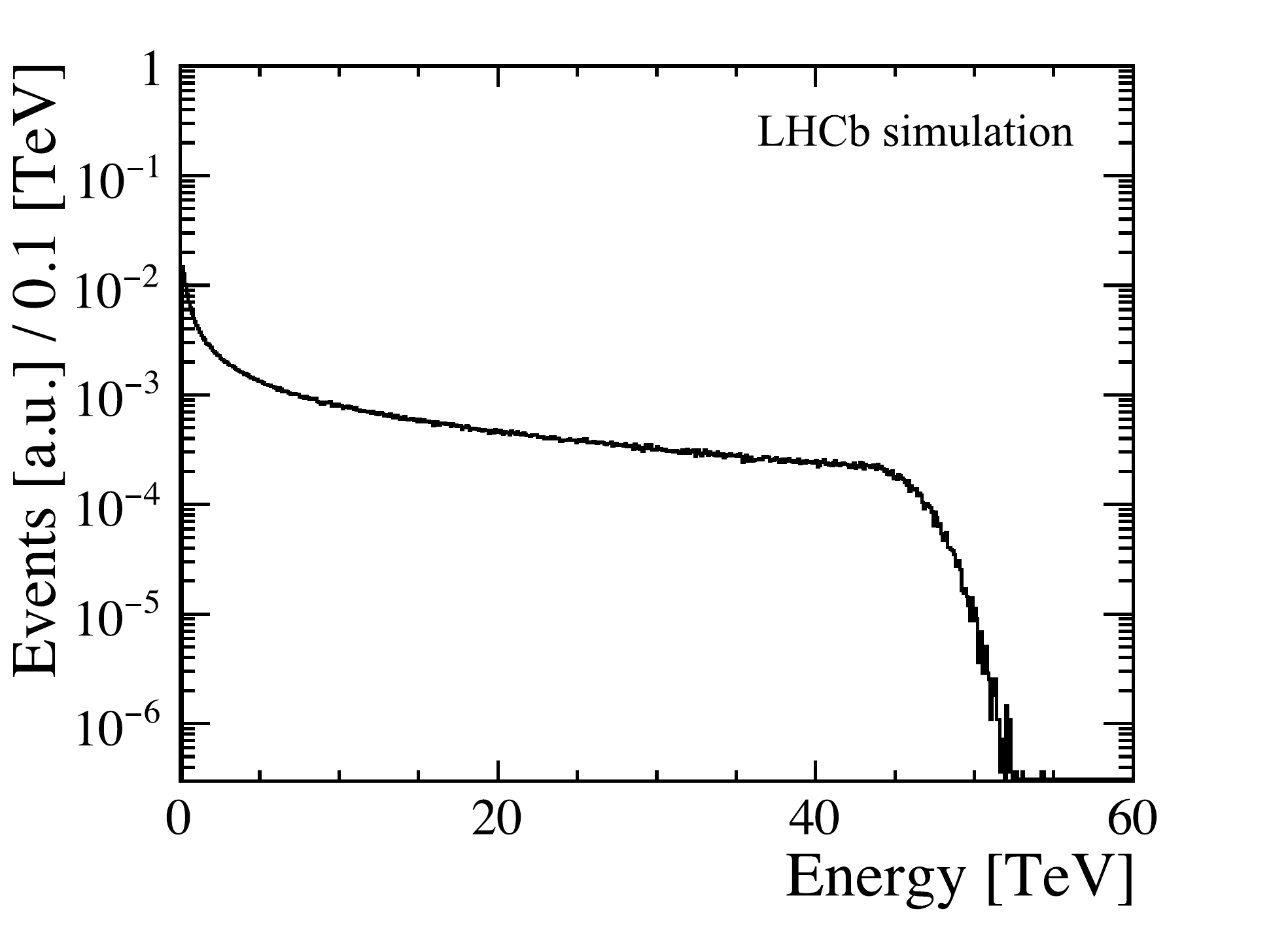}
    \caption{Simulated distribution of the energy deposited in the \ecal for PbPb collisions. A mean energy deposition per particle of $\langle E^{\mathrm{PbPb}} \rangle = 10.4\gev$ is considered.}
    \label{fig:simu_ecal}
  \end{figure}

 The mean energy per particle in the \ecal is assumed to come mainly from $\pi^0$ decays. The $\pi^0$ spectrum is approximated with the charged pion spectrum seen in data in MB $pp$ collisions at 5\tev. The value for the mean energy deposited per particle in the \ecal is found to be $\langle E^{\mathrm{PbPb}} \rangle = 10.4\gev$. The simulated \ecal energy distribution can be seen in Fig.~\ref{fig:simu_ecal}.
  
  The same procedure is repeated for the PbNe case. However, since there are no $pp$ data at \mbox{$\sqsnn=69\gev$}, $p$Ne collisions from the 2015 data-taking period at this centre-of-mass energy are used. The value for the mean energy found is $\langle E^{\mathrm{PbNe}} \rangle = 10.4\gev$. This value is consistent with the one found for PbPb, but this is compensated by a much lower number of particles produced in the collisions.
  
  \subsubsection*{Fit model}
  
\begin{figure}[tbp]
  \centering
  \includegraphics[width=0.49\textwidth]{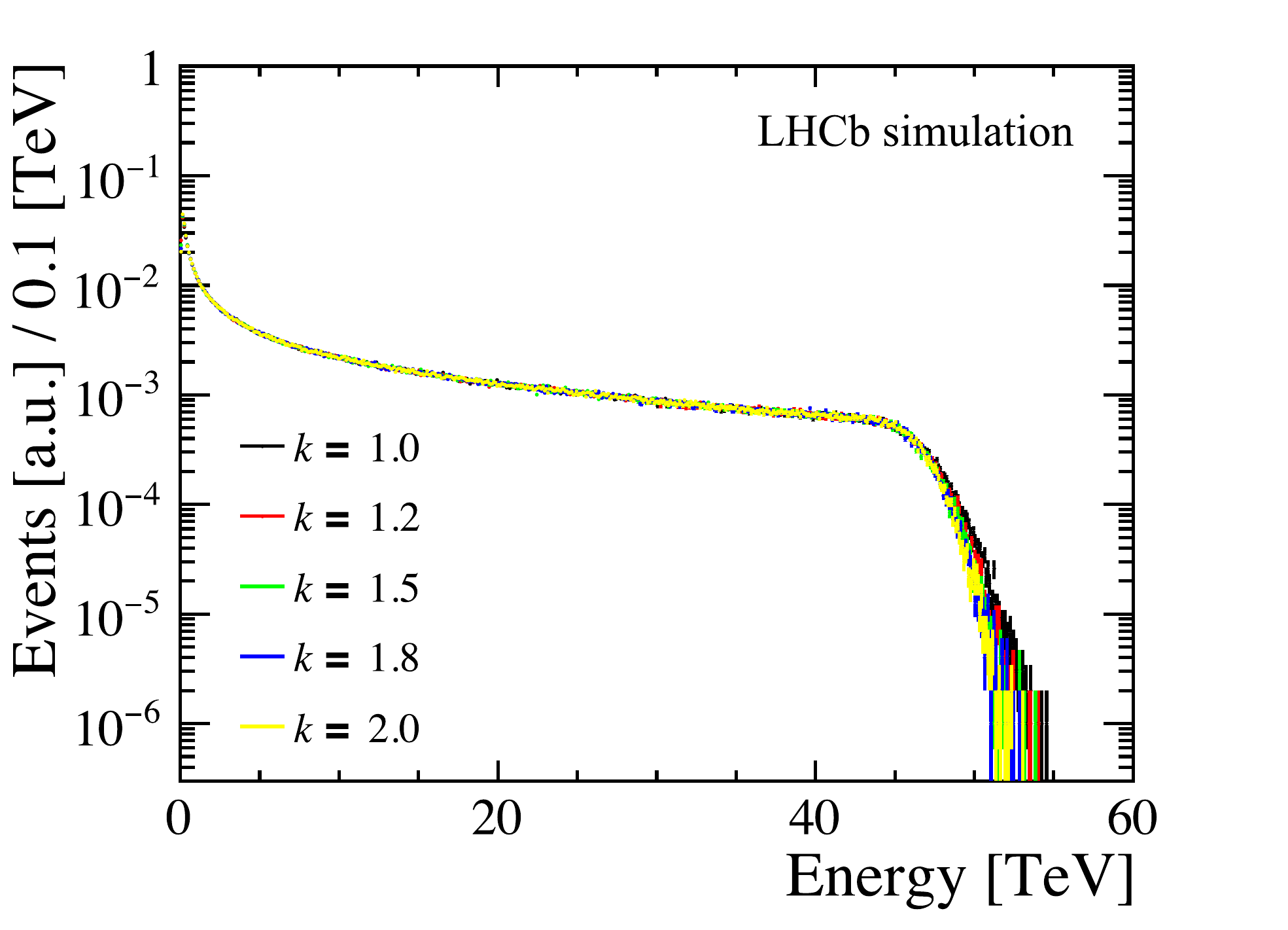}
  \caption{Resulting PbPb simulated energy distribution in the \ecal for $k\in [1.0,\,2.0]$.}
  \label{fig:k_sweep}
  \end{figure}

 The parameter $k$ is linked to the width of the NBD distribution. The ALICE collaboration uses a value $k=1.6$~\cite{Abelev:2013qoq} in their analysis. A comparison is made between PbPb simulated distributions varying $k$ between 1.0 and 2.0 while everything else is kept constant. It is found that there is no significant dependence on this parameter. The resulting distributions can be seen in Fig.~\ref{fig:k_sweep}. For this reason the parameter $k$ was fixed to $k=1.5$ to be in the middle of the explored range.
  Finally the model has only two free parameters, $f$ from Eq.~\ref{eq:nanc} and $\mu$ from the NBD.
  
  \subsubsection*{Fit to the PbPb collision data}
  \label{seq:fittodata}
  
  To fit the simulated distribution to the data a \chisq function on both distributions is minimised. It is defined as
  \begin{equation}
  \chisq = \sum_{i\, \in \,\mathrm{bins}} \frac{(E_i-O_i)^2}{E_i},
  \label{chisq}
  \end{equation}
  where $E_i$ and $O_i$ are the expected and observed values for the $i^{th}$ bin, \ie the simulated and measured values for a given energy bin, assuming the values are counts with Poissonian errors. As previously mentioned, in order to avoid a possible contamination at low energy of electromagnetic origin, the fitting range is chosen to be from 2 to 52\tev. The MC Glauber energy distribution is normalised to the data in the energy range of 5 to 15\tev to avoid the extremes of the distributions.

The fit procedure is delicate since the variables $f$ and $\mu$ are highly correlated as they both modulate the horizontal reach of the distribution: $\mu$ is related to the mean value of the NBD, and thus the high $\mu$ then the higher the number of particles produced per collision and consequently the more energy is deposited in the \ecal. On the other hand, $f$ controls how alike the final distribution is to the distributions of \npart or \ncoll, which have a different reach on the $x$-axis. For the same reason $f$ controls the shape of the right shoulder, which helps disentangle $f$ from $\mu$.

\begin{figure}
  \centering
  \includegraphics[width=0.49\textwidth]{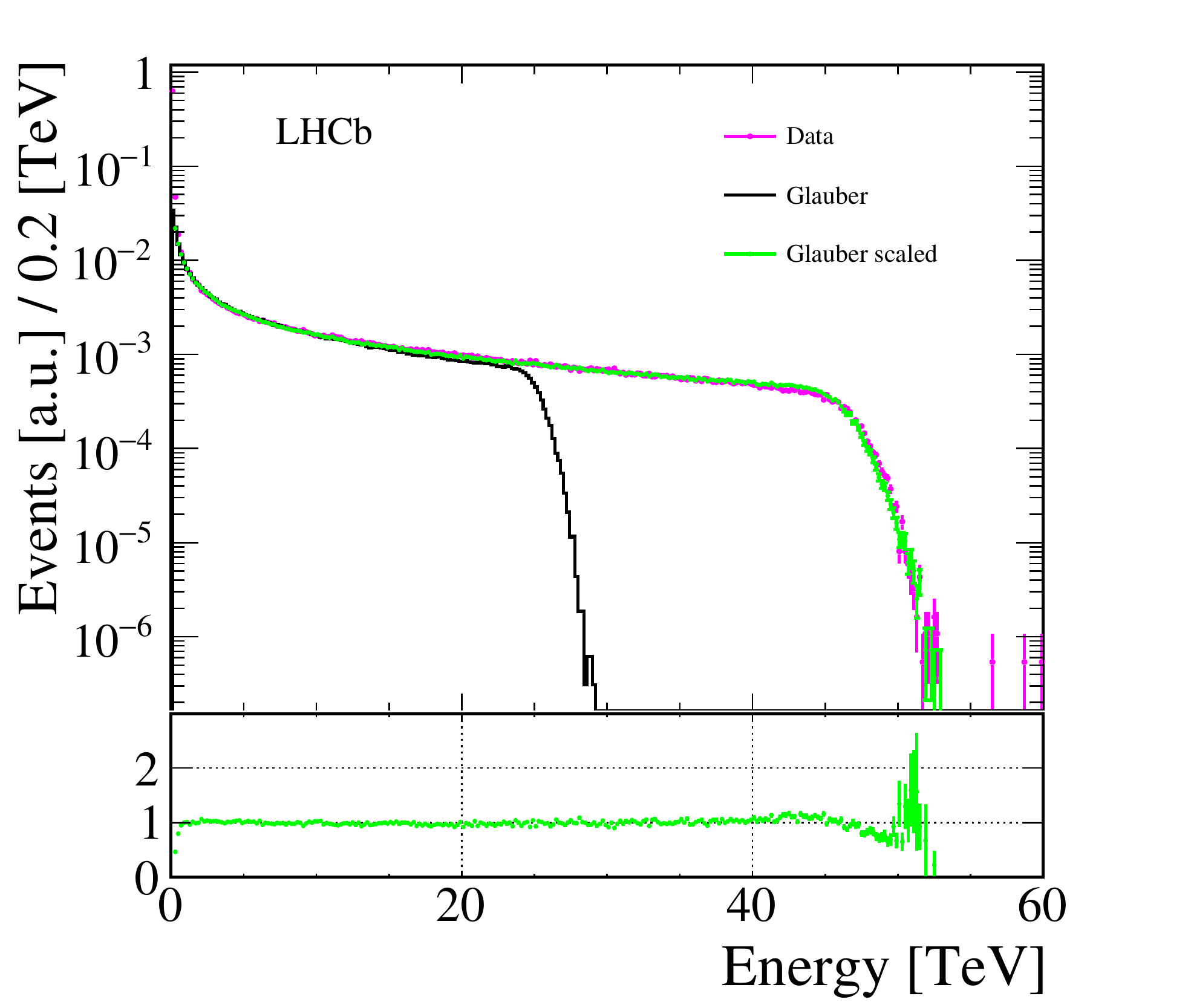}
  \caption{\ecal energy distribution comparison between PbPb data and MC glauber. The black histogram corresponds to the simulated distribution with $f=0.9$ which is then rescaled by $H_s$ (green dots) to match the data and compare the right shoulders. For the entire process $\mu$ was fixed to 3.85.}
  \label{fig:hor_scaling}
  \end{figure}

\begin{figure}
  \centering
  \includegraphics[width=0.49\textwidth]{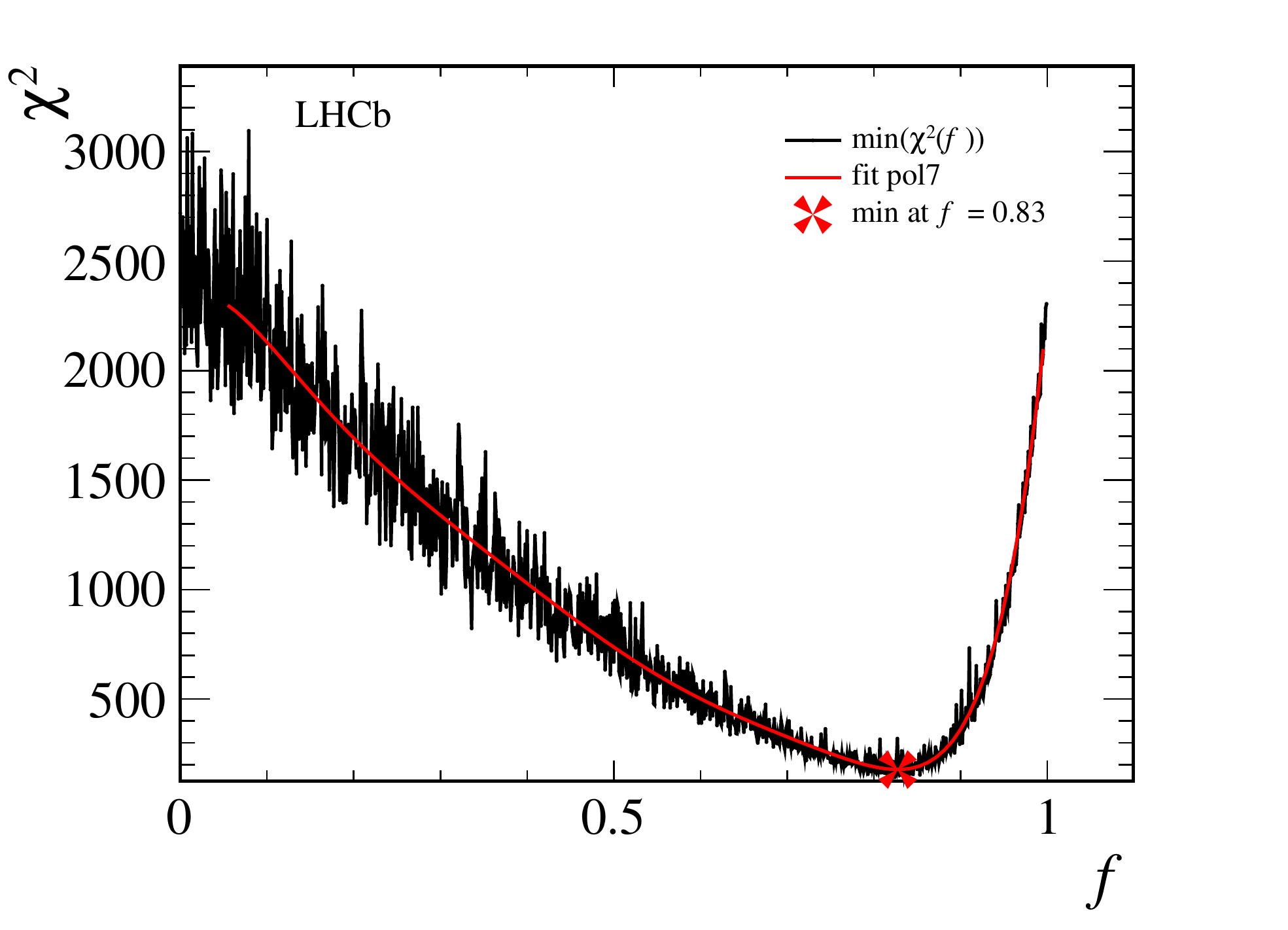}
  \caption{The \chisq values for 1000 steps in $f\in[0,1]$ for the PbPb case. The \chisq values have been fitted by a 7$^{th}$ degree polynomial whose minimum is at $f=0.83$.}
  \label{fig:fsweep_chisq}
  \end{figure}

In the first step $\mu$ is fixed to a test value, $\mu=3.85$, and the similarity between the right shoulder of the data and the simulated distributions is evaluated for 1000 different values of $f$ ranging from 0 to 1. For this procedure the simulated distribution is horizontally scaled to match the data by a factor $H_s$ defined as
\begin{equation}
H_s = \frac{E_{300}^{\mathrm{Data}}}{E_{300}^{\mathrm{Glauber}}},
\end{equation}
where $E_{300}^{\mathrm{Data}}$ is the energy of the last bin containing more than 300 events in the data (bin centre), and $E_{300}^{\mathrm{Glauber}}$ is the same in the simulated distribution. This scaling procedure is shown in Fig.~\ref{fig:hor_scaling}, where the data have been normalised to 1, meaning that 300 events now correspond to roughly $1.6\times10^{-4}$. For this step, the \chisq is computed from 35 to 52\tev to only consider the right shoulder. The resulting values for the \chisq as a function of $f$ can be seen in Fig.~\ref{fig:fsweep_chisq}.

\begin{figure}
  \centering
  \includegraphics[width=0.49\textwidth]{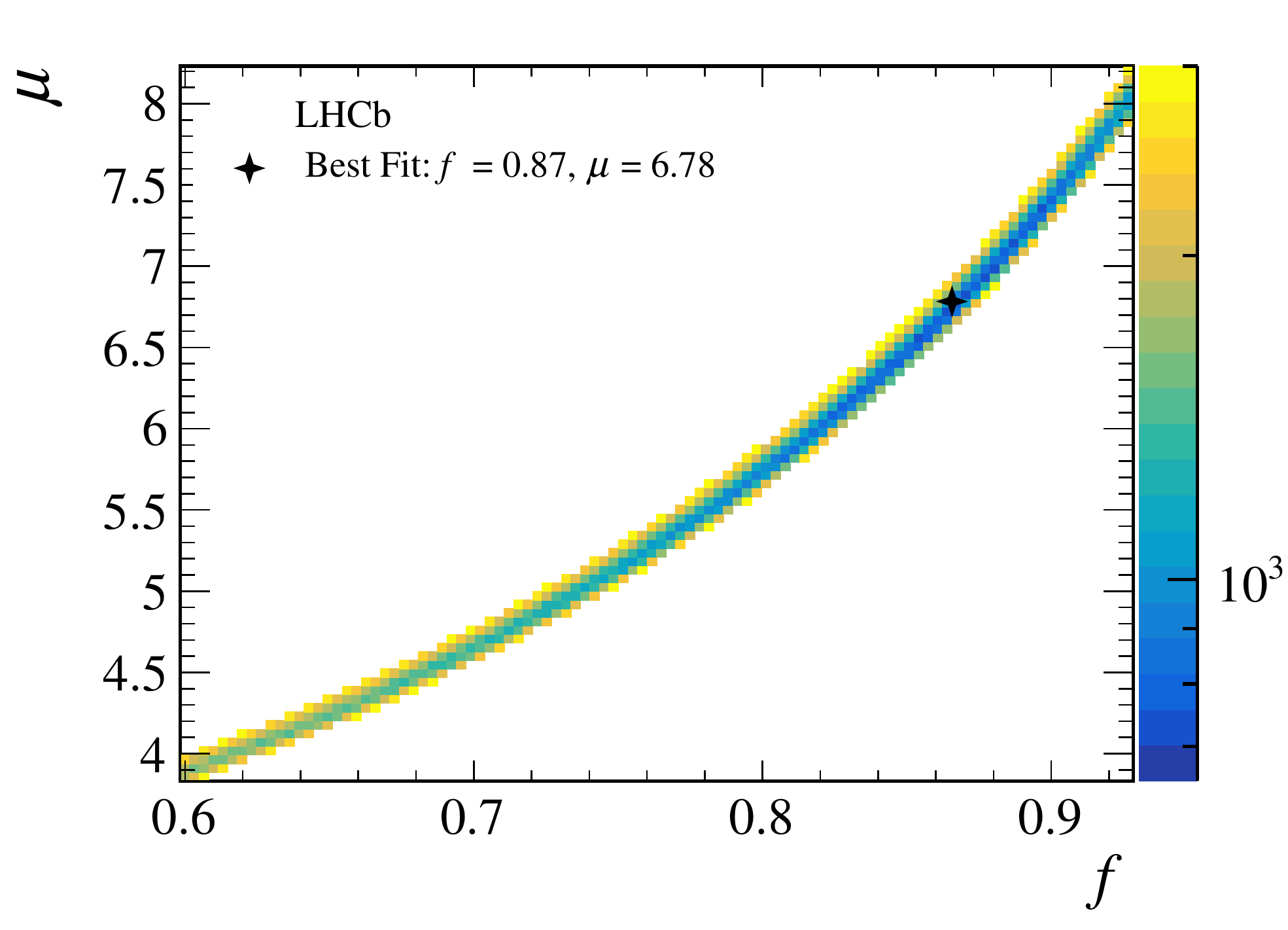}
  \caption{The \chisq map for the coarse grid search in $f\in[0.60,0.93]$ and $\mu\in[3.7,9.0]$ for the PbPb case. The best fit corresponds to the values $f=0.866$ and $\mu=6.778$.}
  \label{fig:chisq_mapcoarse}
  \end{figure}

The minimum of the \chisq is found to be at $f=0.83$. This result is taken as a reference to reduce the range in $f$ for the subsequent grid search. The allowed range for $f$ is set to be $[0.60,0.93]$, from which the range for $\mu$ is chosen to be $[3.7,9.0]$.
A grid of $100\times100$ is defined and the \chisq is computed from 2 to 52\tev at every point of the grid. The result of this grid search can be seen in Fig.~\ref{fig:chisq_mapcoarse}.

\begin{figure}
  \centering
  \includegraphics[width=0.49\textwidth]{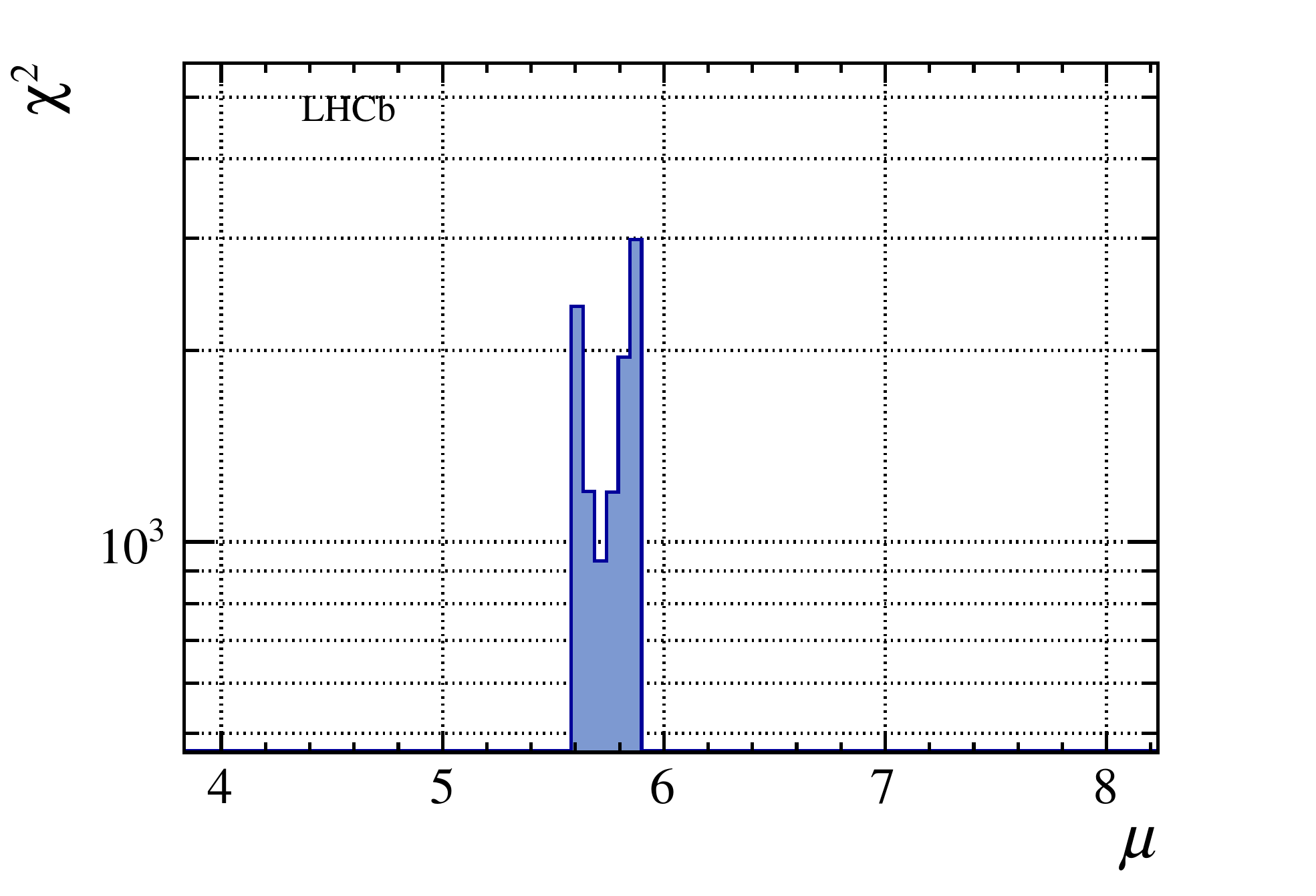}
  \includegraphics[width=0.49\textwidth]{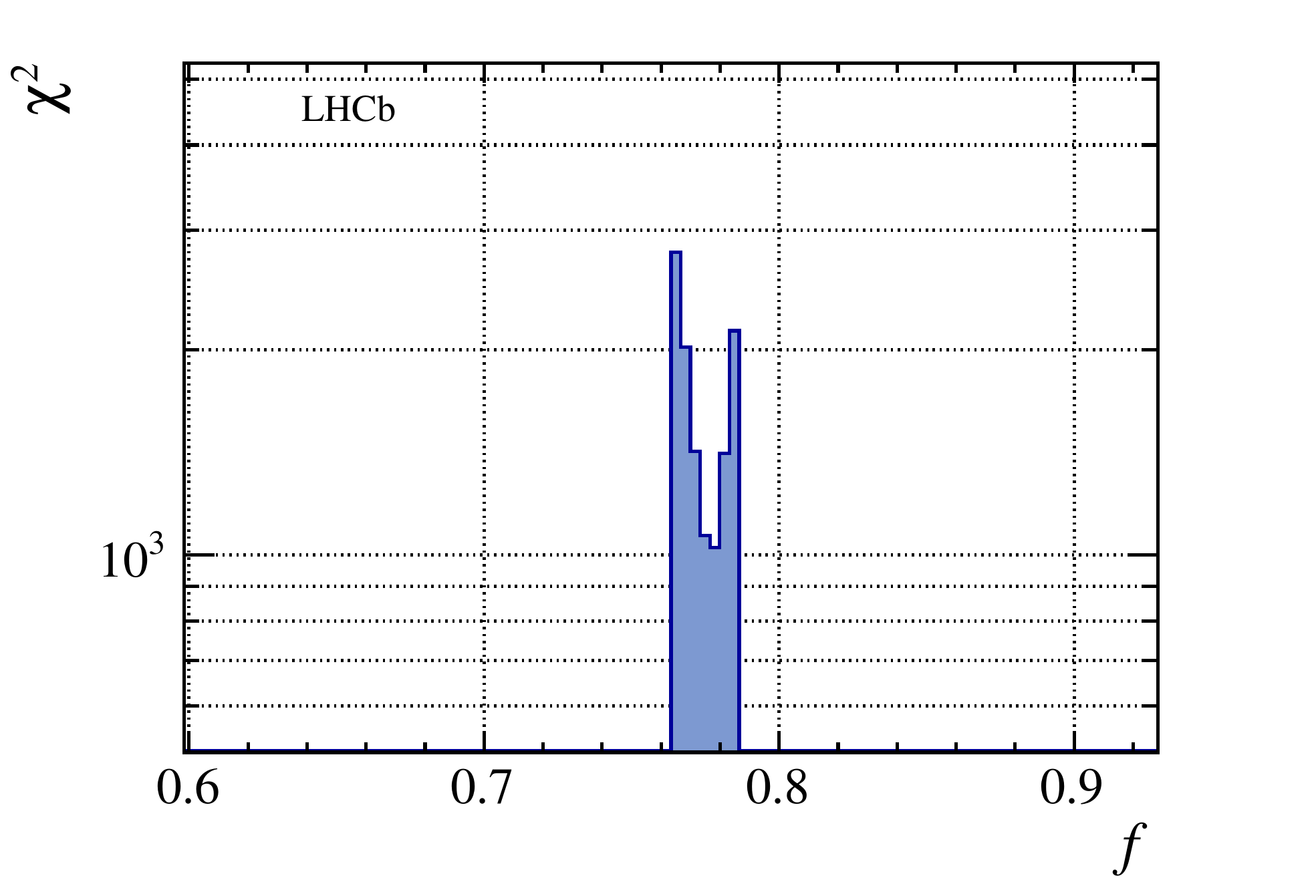}
  \includegraphics[width=0.49\textwidth]{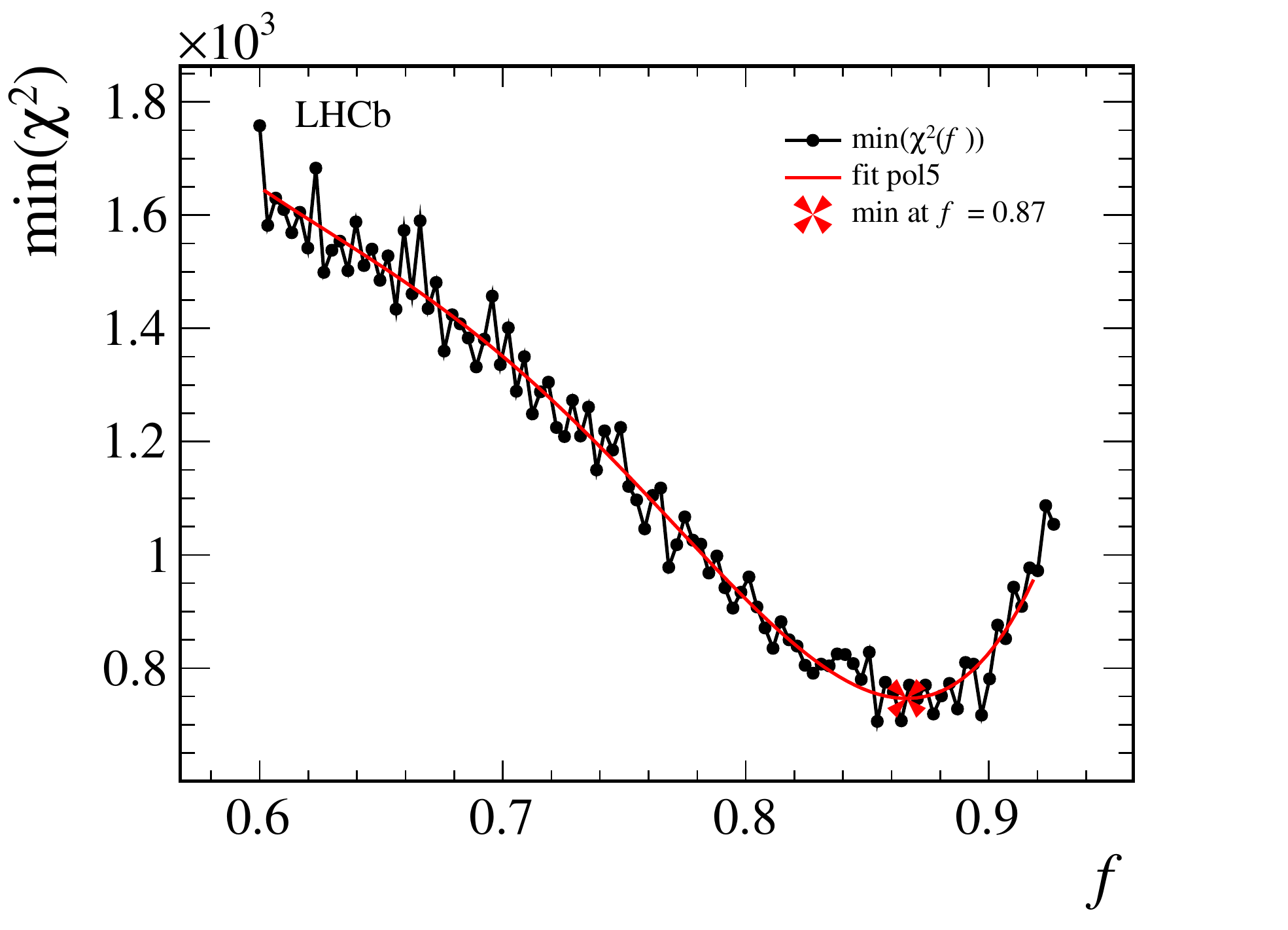}
  \includegraphics[width=0.49\textwidth]{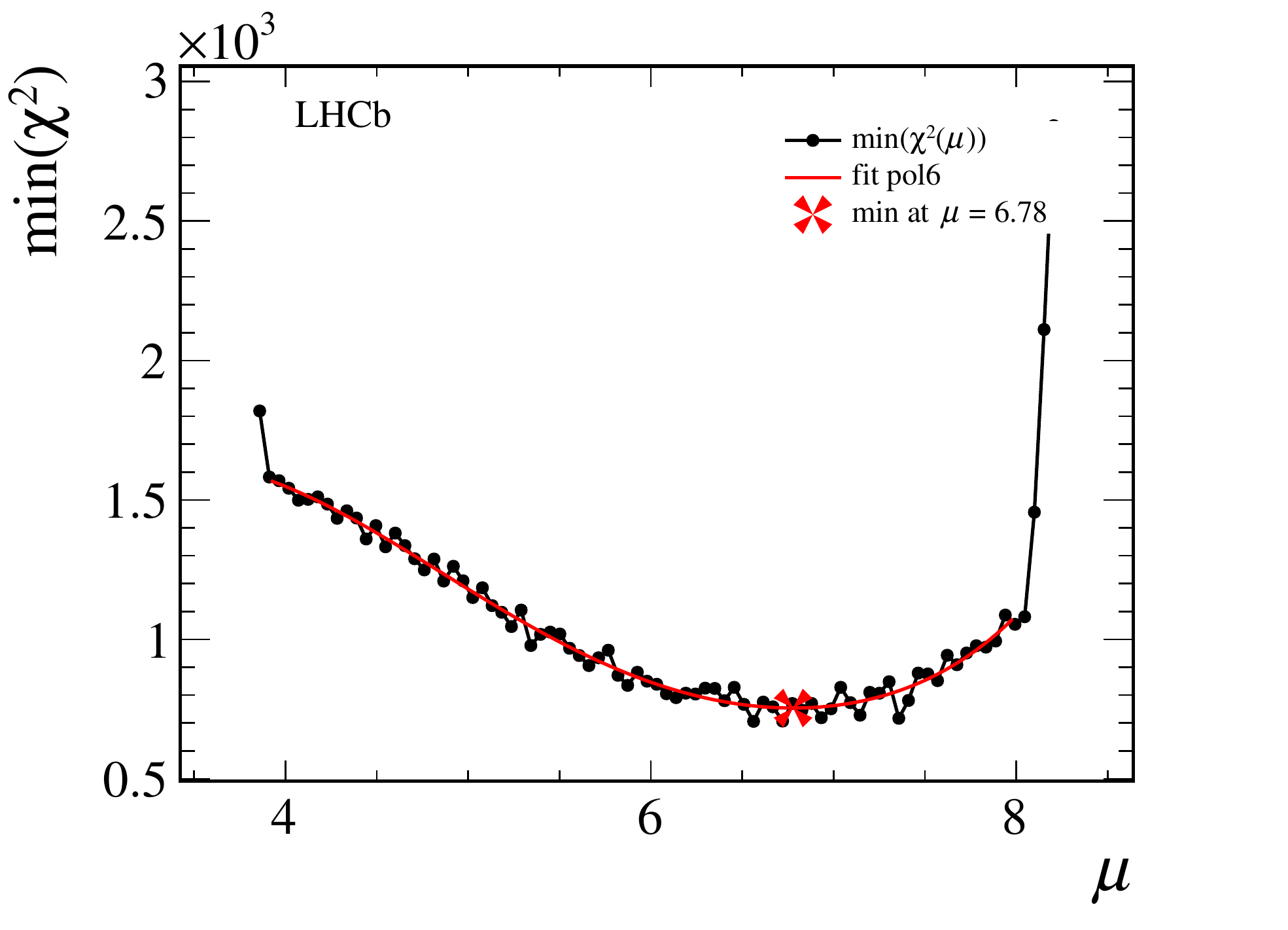}
  \caption{The top plots show a slice for (left) $f=0.798$ and for (right) $\mu=5.45$. From each slice, the minimum of the histogram is kept. The bottom plots show the result of doing this for all values of $f$ and $\mu$, that is, the $f$-parametrised minimum (left) and the $\mu$-parametrised minimum (right) for the PbPb case using the coarse grid. These are fitted by a $5^{th}$ and $6^{th}$ degree polynomial respectively whose minima are at $f=0.866$ and $\mu=6.778$.}
  \label{fig:chisq_parametrisations}
  \end{figure}

  The best fit is not associated to the point of the grid with the lowest value, since this is prone to be affected by the fluctuations from the random NBD sampling. Instead the \chisq map from Fig.~\ref{fig:chisq_mapcoarse} is considered and the minimum parametrised by $f$ and by $\mu$ separately. In order to get the parametrisation as a function of $f$, the minimum of the \chisq as a function of $\mu$ in bins of $f$ is found. Like this, the value of the minimum in each slice along $\mu$ (at fixed $f$) is assigned to the corresponding value of $f$. The same is done for the $\mu$ dependence. Consequently, for all values of $f$ and $\mu$ the \mbox{$f$-parametrised} minimum and \mbox{$\mu$-parametrised} minimum are
  constructed. In Fig.~\ref{fig:chisq_parametrisations} the parametrisations as a function of $f$ and $\mu$ are shown and an example slice is displayed to illustrate the process. From this procedure,
  the best fit is found at $(f,\mu)=(0.866,6.778)$ with a $\chisqndf = 3.025$, which is the one shown in Fig.~\ref{fig:chisq_mapcoarse}. The number of points on the grid is not limiting the precision on the result.

\begin{figure}
  \centering
  \includegraphics[width=0.49\textwidth]{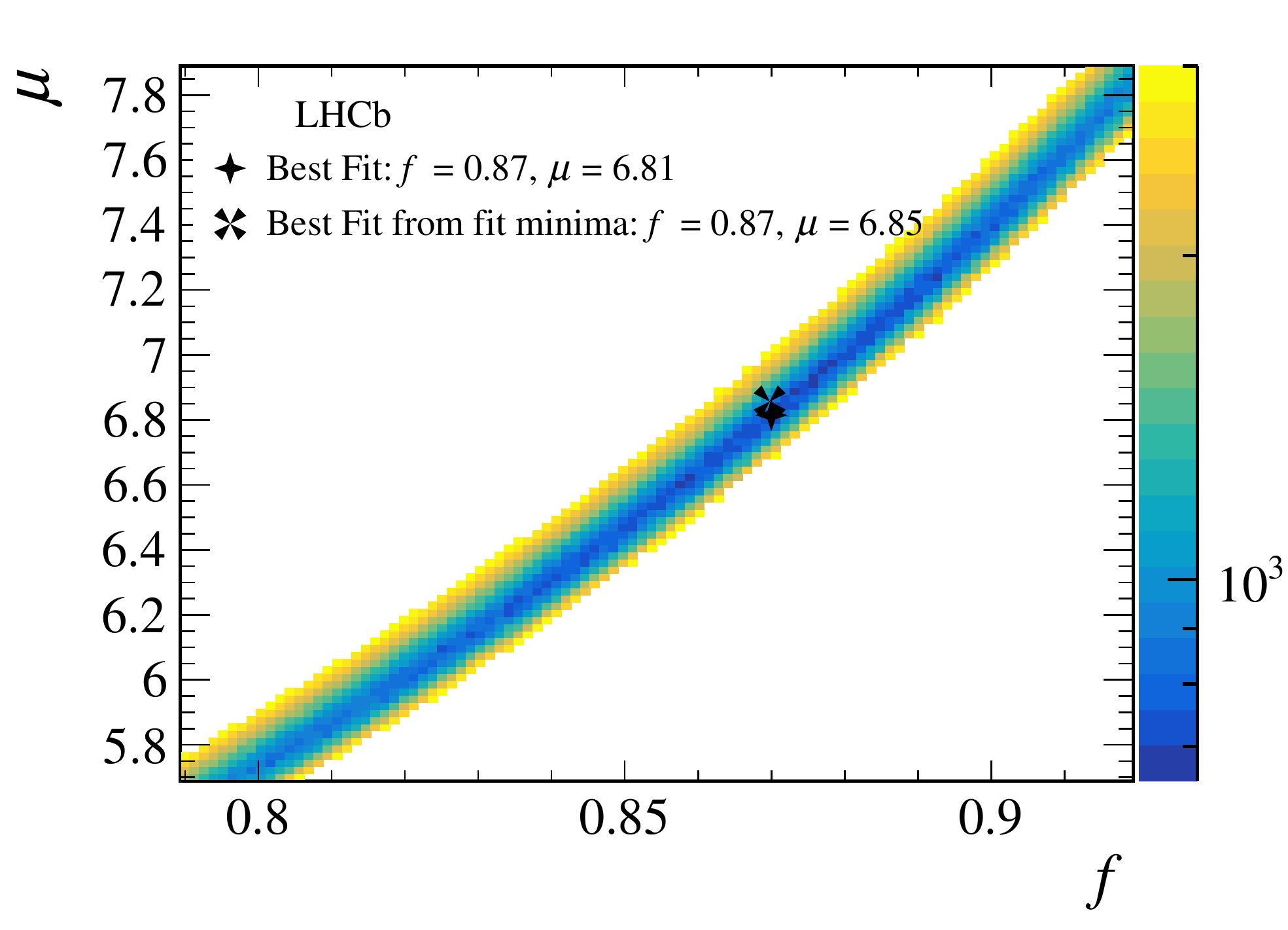}
  \caption{Map of \chisq values for the fine grid search in $f\in[0.79,0.92]$ and $\mu\in[5.7,7.9]$ for the PbPb case. The two shown best fits correspond to the results from two different methods (see text).}
  \label{fig:chisq_mapfine}
  \end{figure}

Finally another grid search is performed with the same amount of points but on a narrower range, namely $f\in[0.79,0.92]$ and $\mu\in[5.7,7.9]$. The resulting \chisq map is shown in Fig.~\ref{fig:chisq_mapfine}. Two results are shown as a best fit. The star is the nominal value, whereas the cross is obtained by using an alternative method where the slice is fitted to obtain the minimum of this fit. The best fits found are $(f,\mu)=(0.869, 6.814)$ with $\chisqndf = 2.82$ and $(f,\mu)=(0.869, 6.853)$ with $\chisqndf = 2.83$. These best fits correspond to the star and the cross on Fig.~\ref{fig:chisq_mapfine} respectively. Since the goodness of fit is virtually the same for the two best fits, the one which is kept is $(f,\mu)=(0.869, 6.814)$.

\begin{figure}[]
  \centering
  \includegraphics[width=0.49\textwidth]{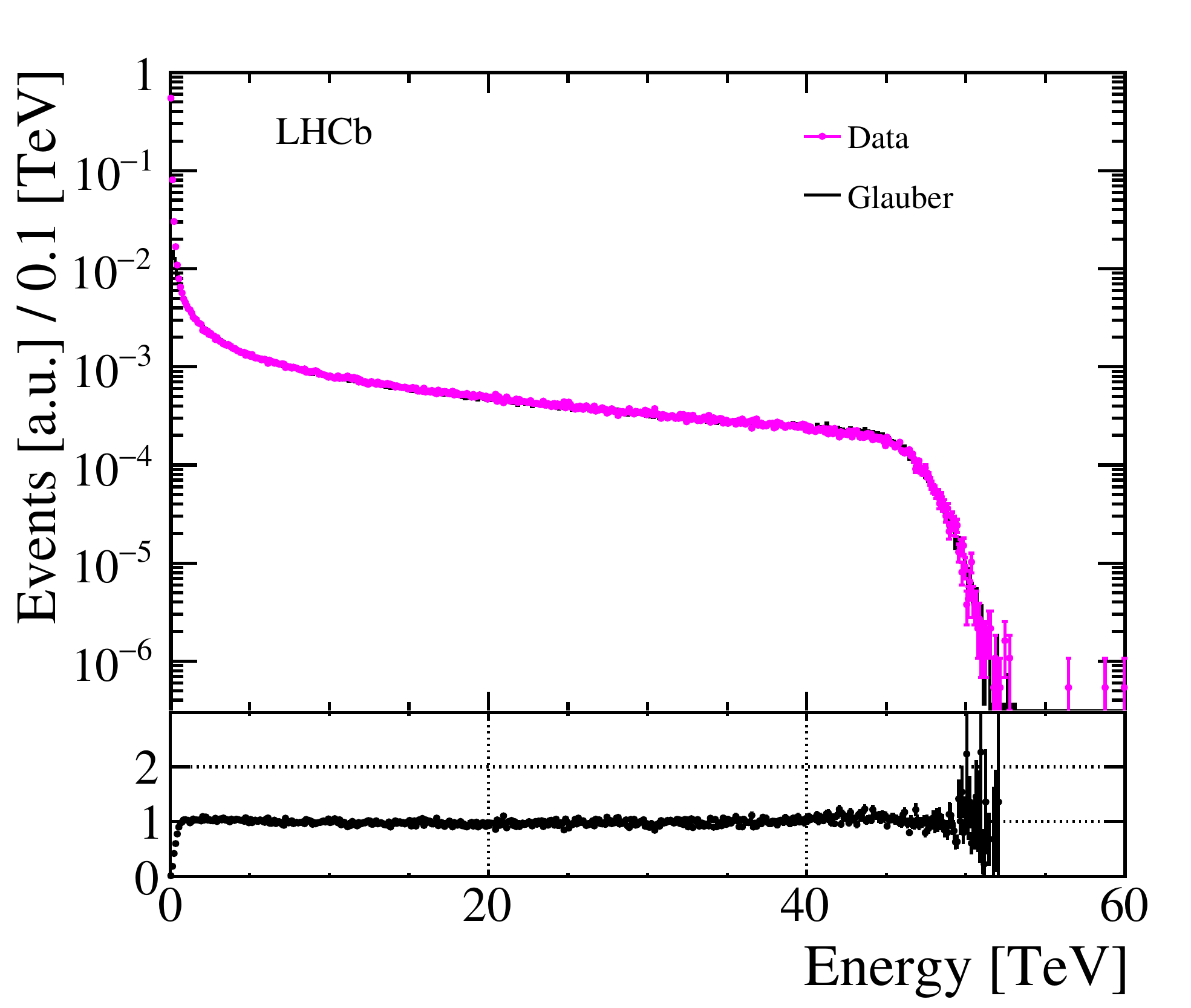}
  \includegraphics[width=0.49\textwidth]{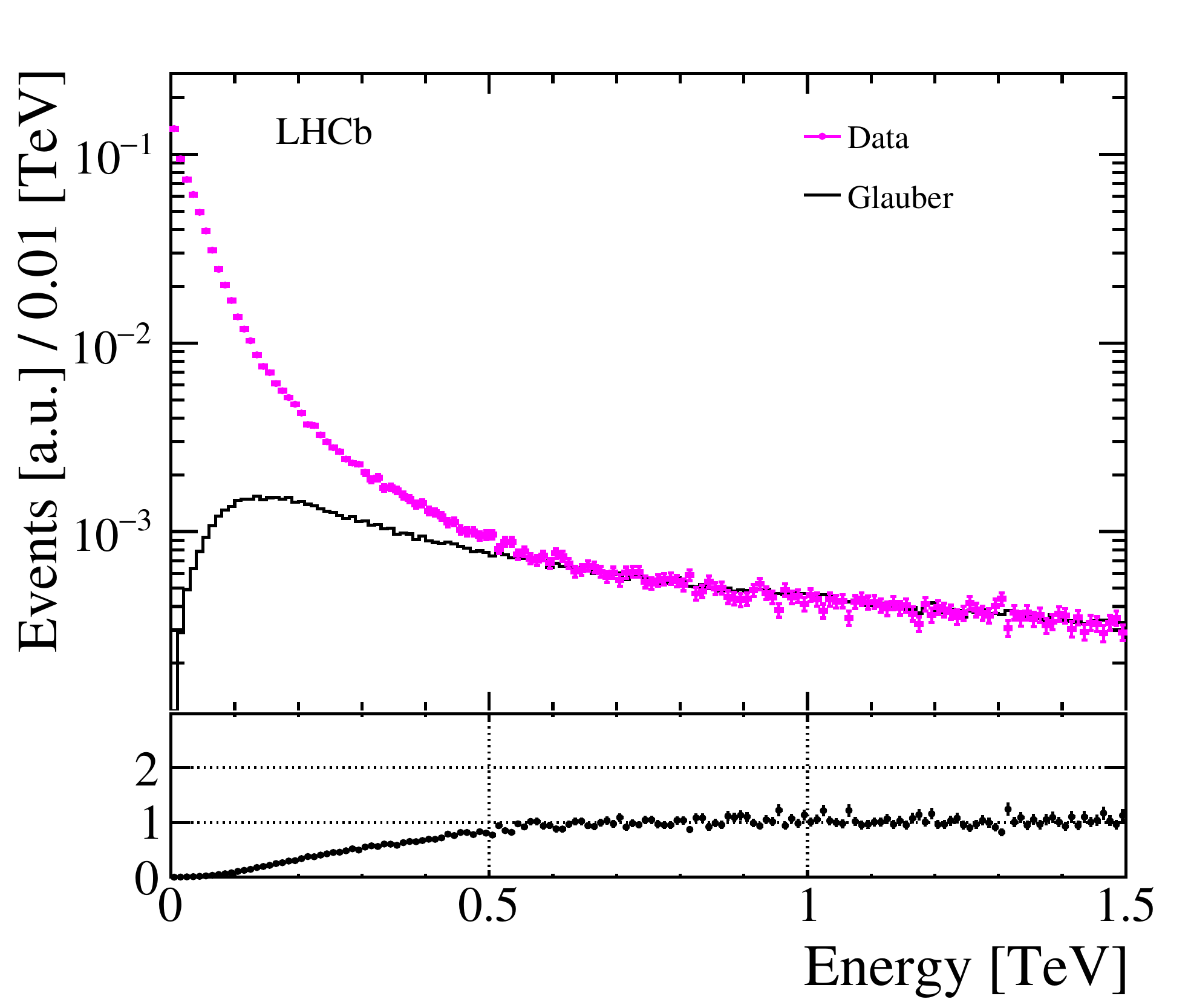}
  \caption{Final fit of the simulated energy distribution to the data for PbPb collisions. The best fit found is $(f,\mu)=(0.869, 6.814)$ with a corresponding $\chisqndf = 2.82$. The right figure corresponds to a close-up view of the left figure.}
  \label{fig:fit_to_data}
  \end{figure}

  The final result of the fit can be seen in Fig.~\ref{fig:fit_to_data}. On the right plot of Fig.~\ref{fig:fit_to_data} a zoom of the low-energy part of the distribution is displayed, where the discrepancy between the MC Glauber and the data, due to the presence of events of electromagnetic origin, becomes clear. This will be addressed in more detail in Sect.~\ref{results}. This region below 0.5\tev is well outside the fitting range, which starts at 2\tev.  The Glauber model, with its parameters obtained from a fit to the data, can thus be used to define the centrality classes in PbPb collision data.  
    
\subsubsection*{Fit to the PbNe collision data}

The same \chisq function from Eq.~\ref{chisq} is used to evaluate the goodness of fit of the Glauber MC to the data for PbNe collisions. The fitting range is chosen to be from 0.5 to 3.9\tev in order to avoid possible contamination from electromagnetic origin present at low energy. The Glauber MC energy distribution is normalised to the data in the energy range of 0.5 to 2\tev to not consider the tails of the distributions, even in the case where $f$ takes on the extreme values.
 
\begin{figure}
  \centering
  \includegraphics[width=0.49\textwidth]{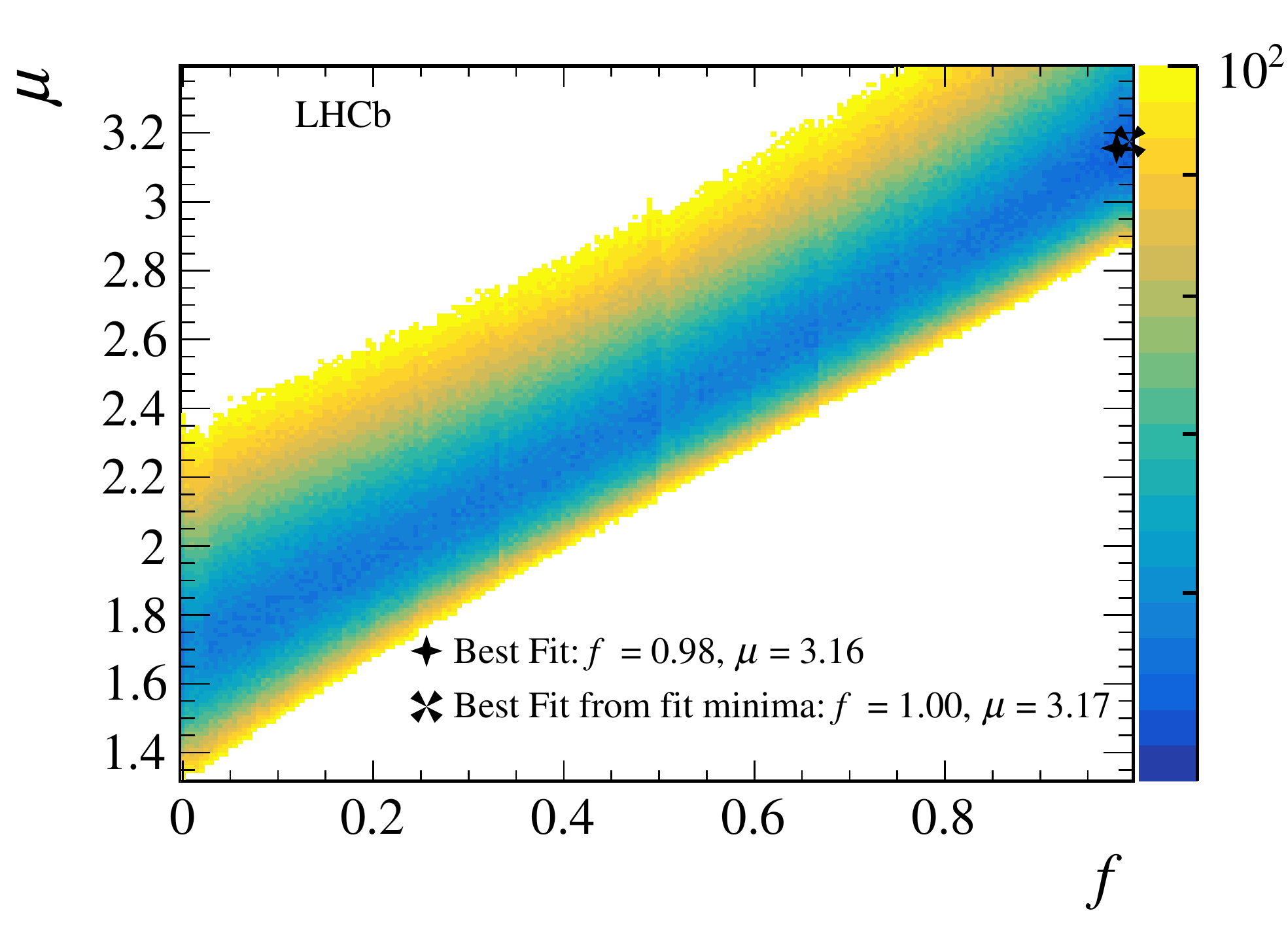}
  \caption{Map of \chisq values for the coarse grid search in $f\in[0.0,1.0]$ and $\mu\in[1.0,3.4]$ for the PbNe case. The two best fits shown correspond to the results of two different methods described in the text.}
  \label{fig:chisq_mapcoarse_pbne}
  \end{figure}

Since in this scenario the shape of the energy distribution at high energy is not as characteristic as it is for PbPb, the approach of trying to fit $f$ first cannot be applied, and the allowed range remains $f\in[0.0,1.0]$, since there is still sensitivity to the contributions of \npart or \ncoll. The range in $\mu$ is chosen accordingly to be $\mu\in[1.0,3.4]$. A grid of $200\times200$ is defined in the previously mentioned ranges and the \chisq is computed at every point of the grid. The result of this grid search can be seen in Fig.~\ref{fig:chisq_mapcoarse_pbne}.

To find the best fits shown in Fig.~\ref{fig:chisq_mapcoarse_pbne}, the same approach described for the PbPb case is used. The values of $f$ and $\mu$-parametrised minima are obtained by taking the minimum value for every slice (first method, star in Fig.~\ref{fig:chisq_mapcoarse_pbne}), and by fitting each slice and getting the minimum of the fit (second method, cross in Fig.~\ref{fig:chisq_mapcoarse_pbne}). However, in the following step only the $\mu$-parametrised minima are fit to get the optimal $\mu$ whereas for the $f$-parametrised minima, the $f$ value where the \chisq is minimum is picked without fit. This is because the $f$-parametrised distributions show a steep decrease as $f$ approaches one, and fitting a function around that region is difficult. The best fit is found at $(f,\mu)=(0.980,3.156)$ with a $\chisqndf = 1.039$, and at $(f,\mu)=(0.995,3.174)$ with a $\chisqndf = 1.031$. These best fits are shown in Fig.~\ref{fig:chisq_mapcoarse_pbne} by a star and a cross, respectively.

\begin{figure}
  \centering
  \includegraphics[width=0.49\textwidth]{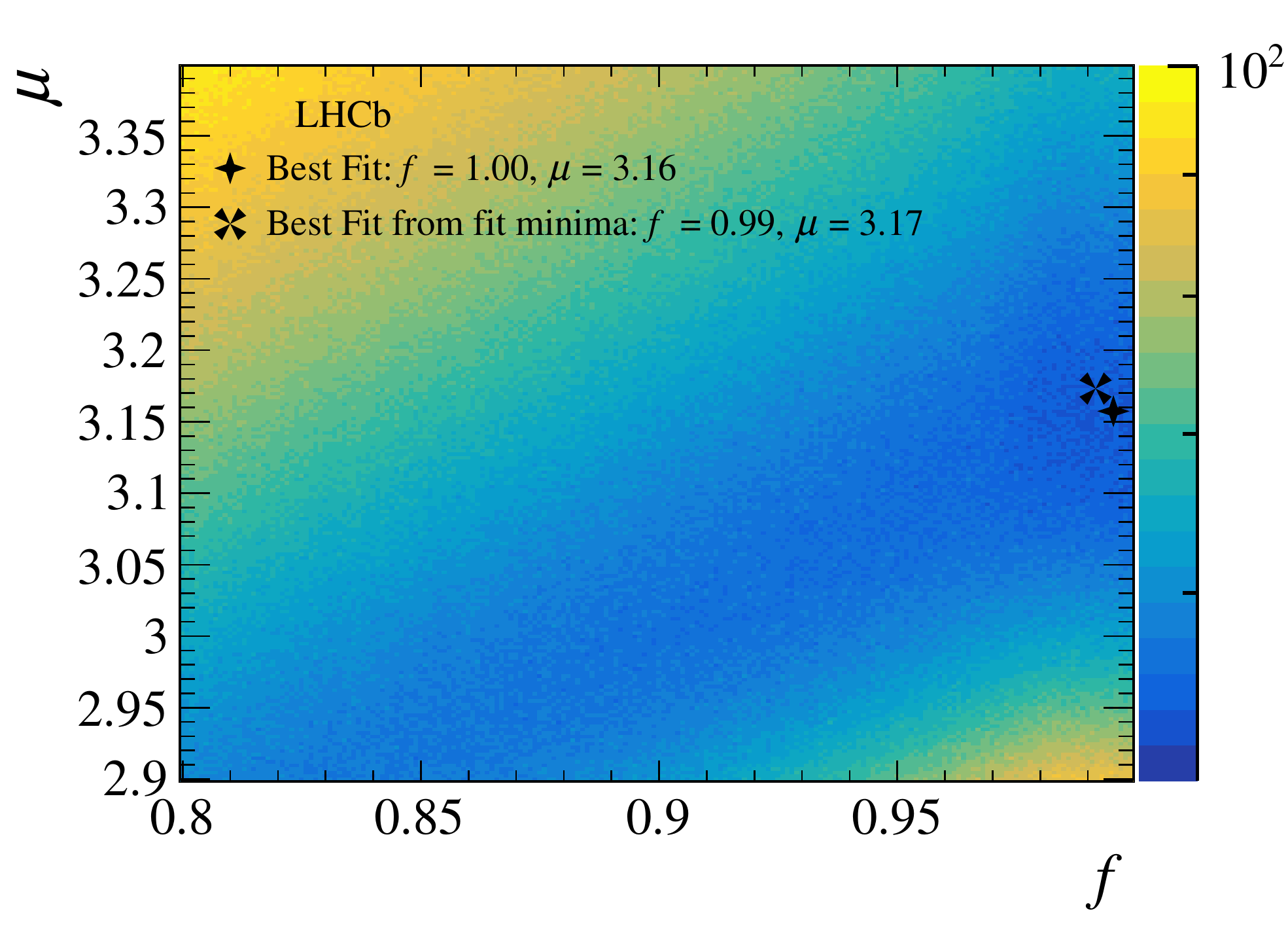}
  \caption{Map of \chisq values for the fine grid search in $f\in[0.8,1.0]$ and $\mu\in[2.9,3.4]$ for the PbNe case. The two shown best fits correspond to the results from two different methods described in the text.}
  \label{fig:chisq_mapfine_pbne}
  \end{figure}

Finally another grid search is performed where the same amount of points is used but on a narrower range, namely $f\in[0.8,1.0]$ and $\mu\in[2.9,3.4]$. From this grid, the best fits are found at $(f,\mu)=(0.996,3.157)$ with a $\chisqndf = 1.026$, and at $(f,\mu)=(0.992,3.173)$ with a $\chisqndf = 1.031$, computed using the same methods described above. The resulting \chisq map and the best fits are shown in Fig.~\ref{fig:chisq_mapfine_pbne} by a star and a cross, respectively. Since the goodness of fit is virtually the same for all best fits in the coarse and fine grid, the one kept is the one which results in the smallest \chisq value, that is, $(f,\mu)=(0.996, 3.157)$.

\begin{figure}
  \centering
  \includegraphics[width=0.49\textwidth]{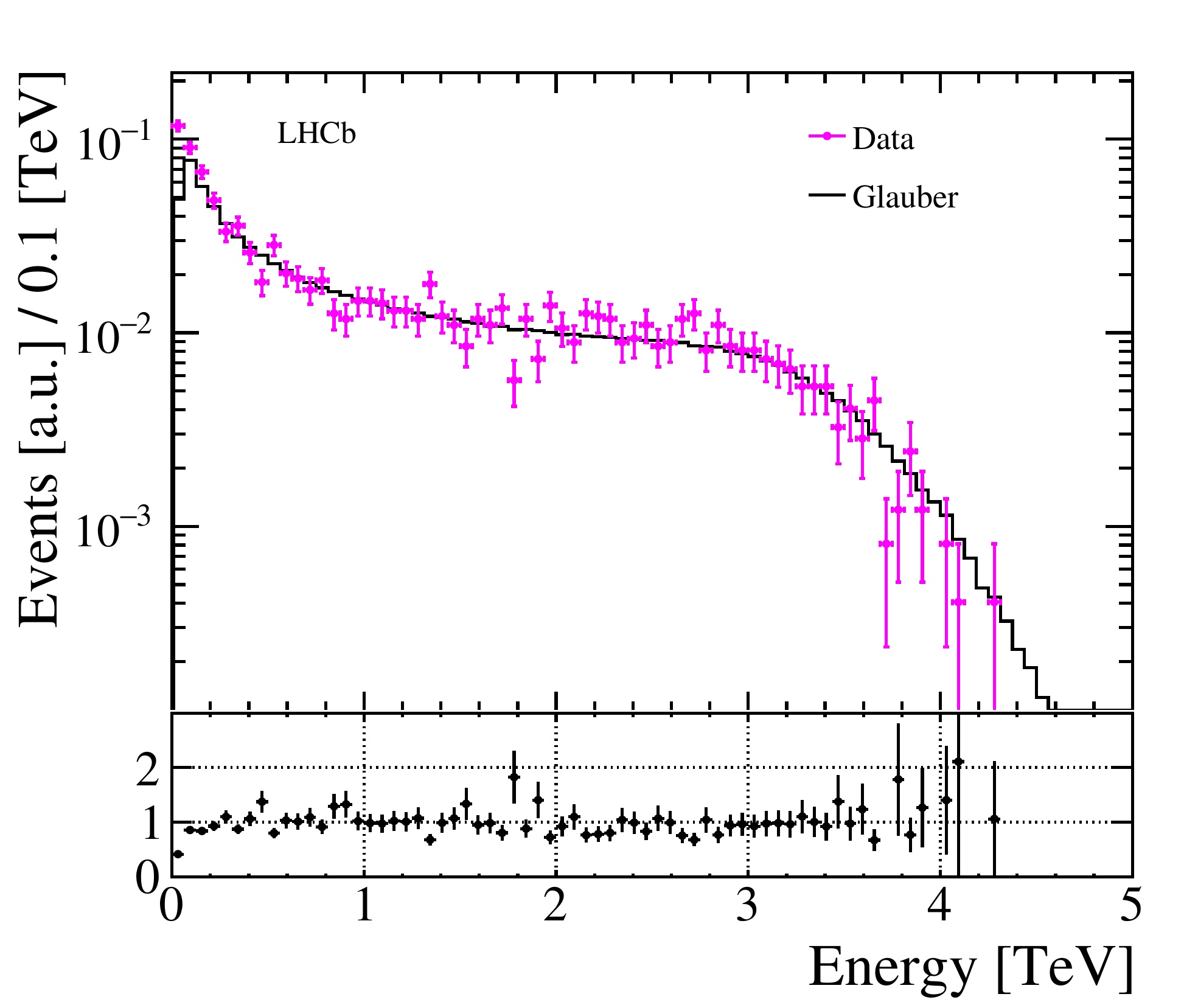}
  \includegraphics[width=0.49\textwidth]{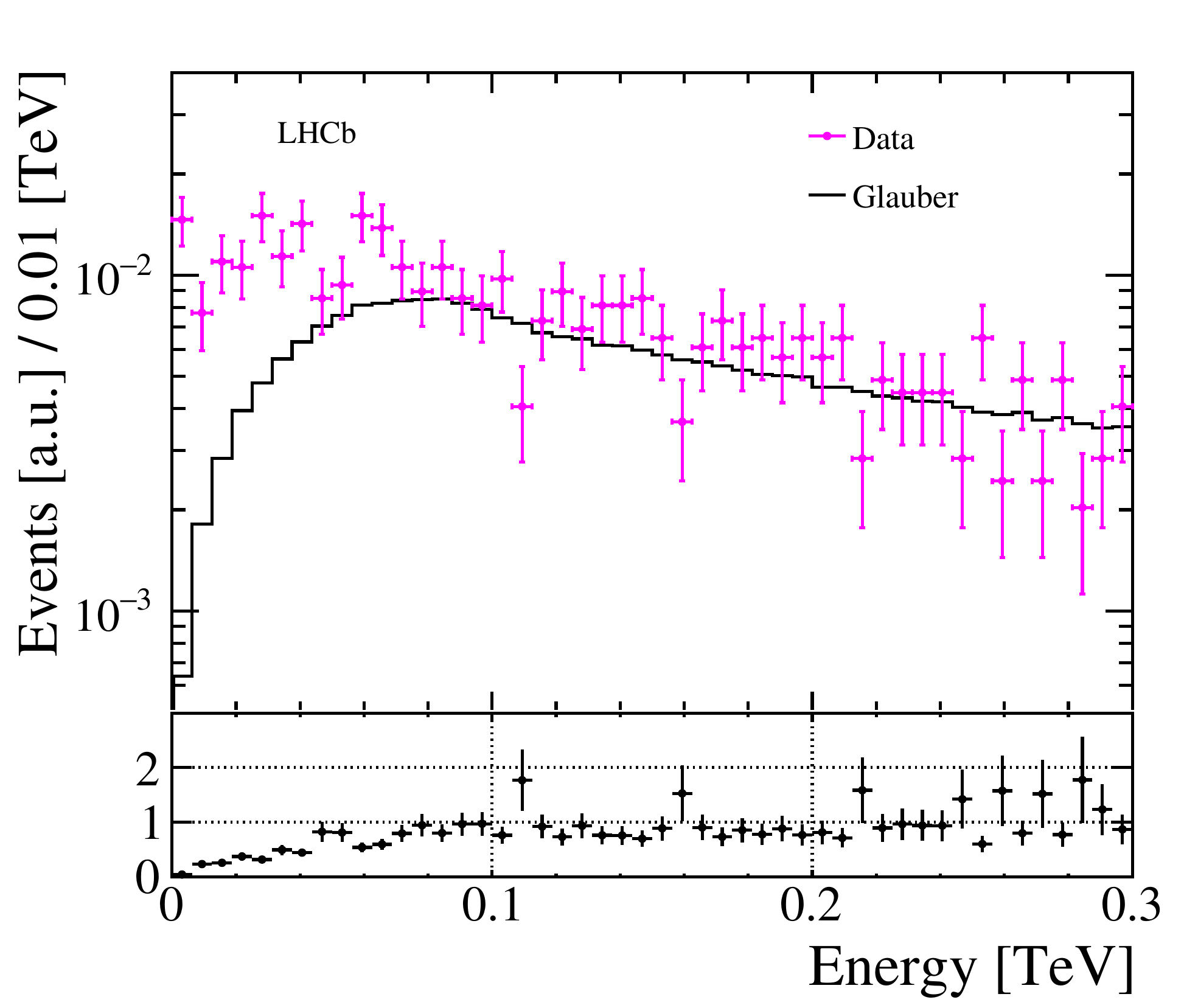}
  \caption{Final fit of the simulated energy distribution to the data for PbNe collisions. The best fit found is $(f,\mu)=(0.996, 3.157)$ with a corresponding $\chisqndf = 1.026$. The right figure corresponds to a close-up view of the left figure.}
  \label{fig:fit_to_data_pbne}
  \end{figure}

The final result of the fit can be seen in Fig.~\ref{fig:fit_to_data_pbne}. On the right plot, a zoom of the low-energy part of the distribution is shown, where the discrepancy between the Glauber MC and the data, due to the presence of events of electromagnetic origin, becomes clear. This region below 0.1\tev is well below the fitting range, \mbox{which starts at 0.5\tev.} 
    
The fact that $f$ is close to one (hence $\nanc\sim\npart$) is due to the fact that, below 100\gev, the particle production is dominated by soft processes, which scale geometrically, as mentioned in Sect.~\ref{sec:sim_evt_pbpb}.
  
\subsubsection*{Centrality classes}
\label{sec:def_cent}

After obtaining the simulated distribution of energy deposited in the \ecal, which corresponds only to the hadronic contribution, the distribution can be divided into centrality classes. To determine the \ecal energy boundary values for each class, the simulated distribution is 
integrated from a value of deposited energy to infinity, until a starting value is found giving a percentage of the total integral. Defining $I_{\text{T}}$ as the total integral of the energy distribution, the \ecal energy requirement for any percentage $p$ of centrality, would be the value of $E_{p}$ such that
\begin{equation}
(p\times 10^{-2})\,I_{\text{T}} = \int_{E_{p}}^{\infty}\frac{\mathrm{dN}}{\mathrm{d}E}\mathrm{d}E.
\end{equation}
Similarly, as an example, the centrality class $(10-20)\%$ would correspond to the events depositing an energy $E$ such that $E_{20}<E<E_{10}$.

\subsection{Results}
\label{results}

\begin{figure}[]
    \centering
      \includegraphics[width=0.48\textwidth]{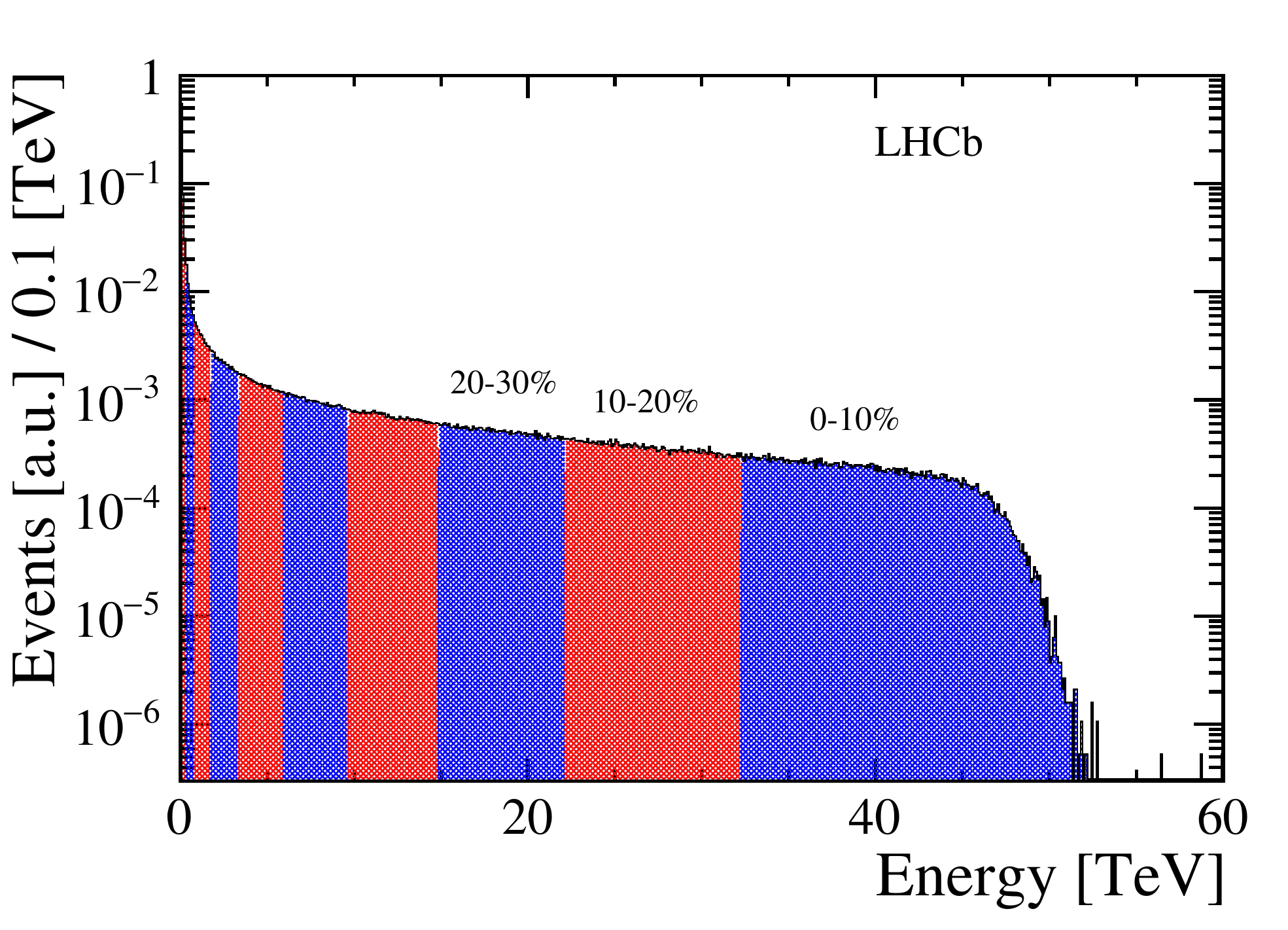}
      \includegraphics[width=0.48\textwidth]{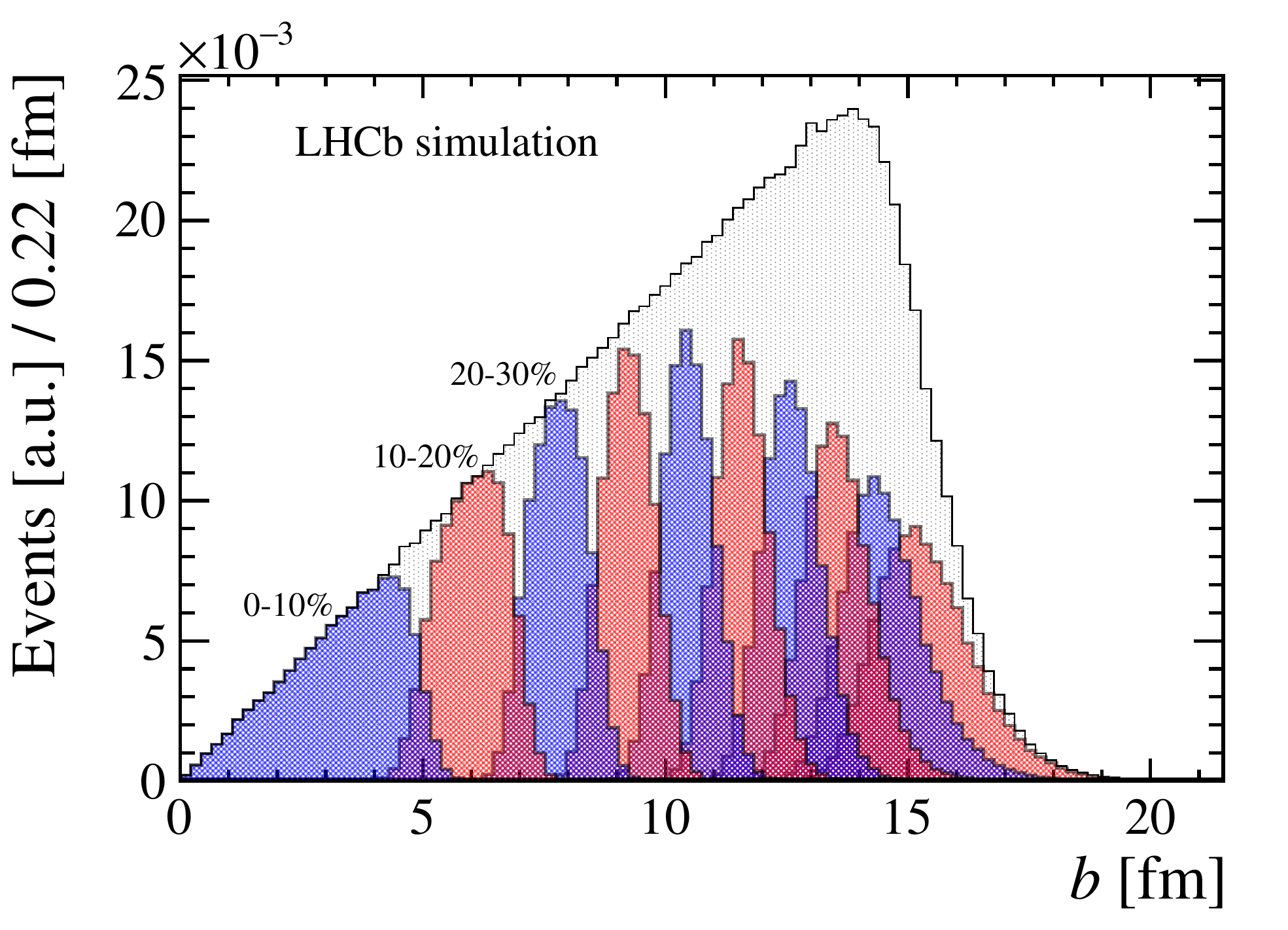}
      \includegraphics[width=0.48\textwidth]{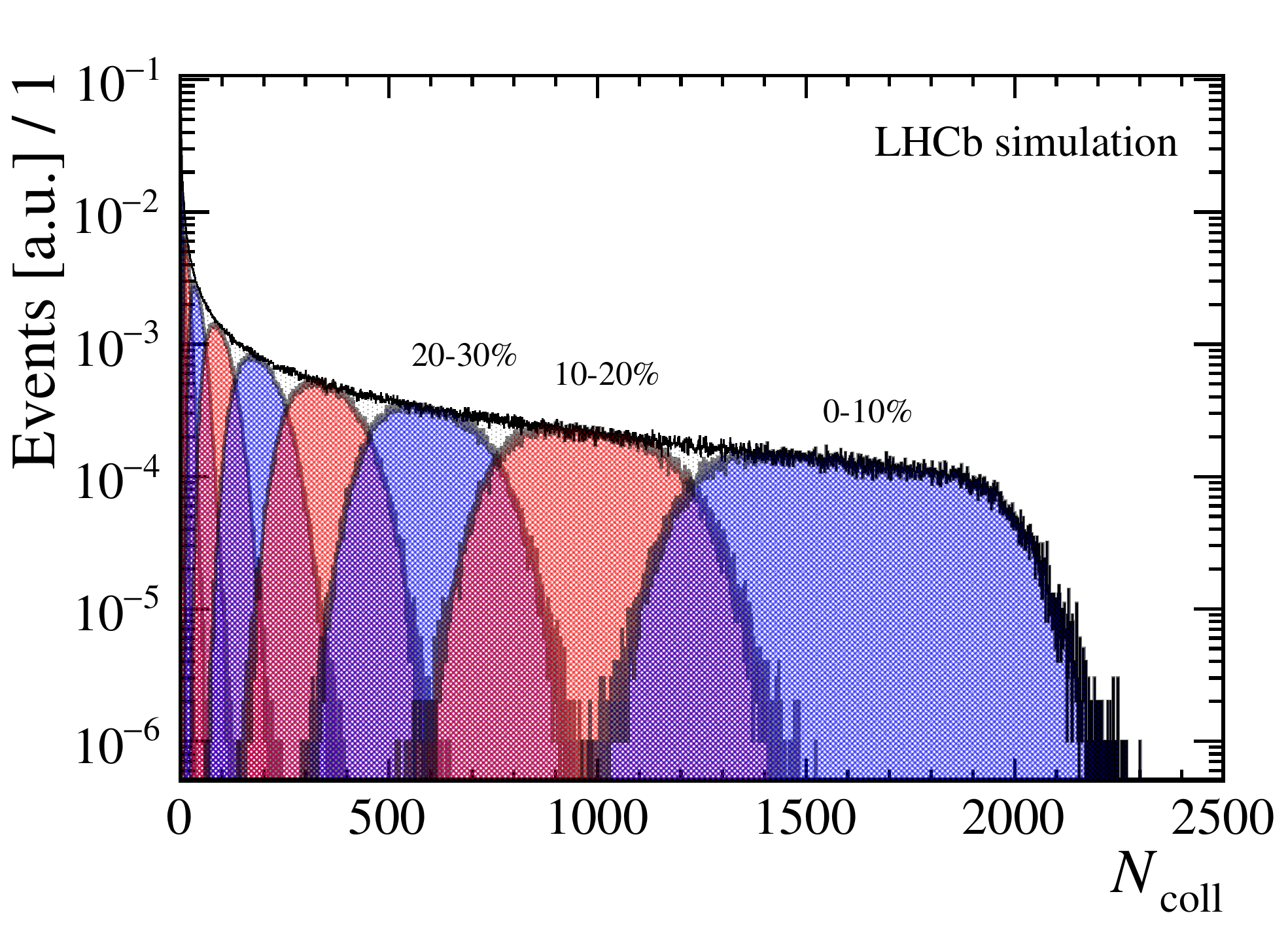}
      \includegraphics[width=0.48\textwidth]{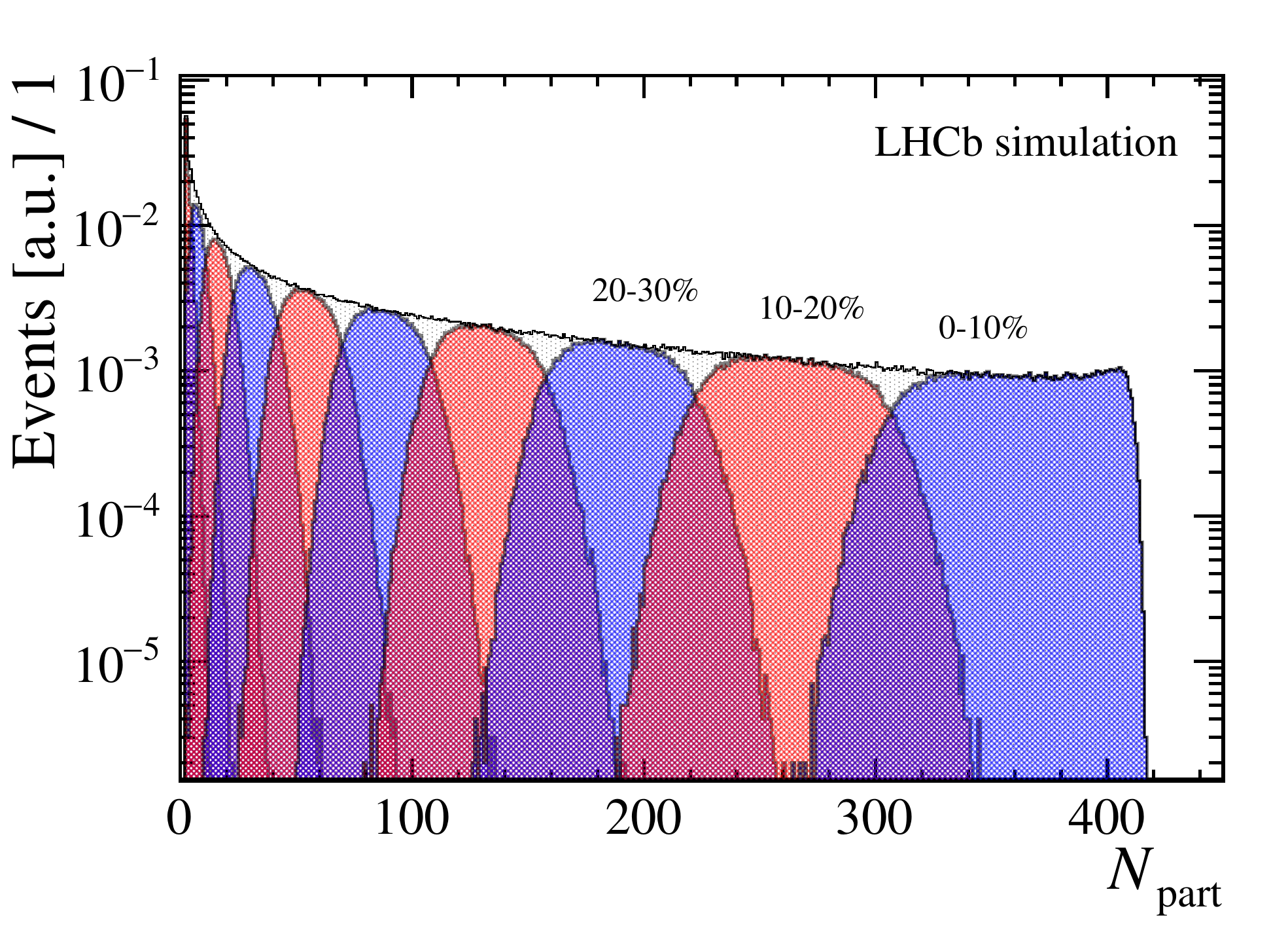}
    \caption{(top left) Classification of events from PbPb data according to the defined centrality classes, distribution of the (top right)  impact parameter, (bottom left) \ncoll and (bottom right) \npart quantities for the corresponding centrality classes.}
    \label{fig:cent_classes}
  \end{figure}

\begin{figure}[]
    \centering
      \includegraphics[width=0.48\textwidth]{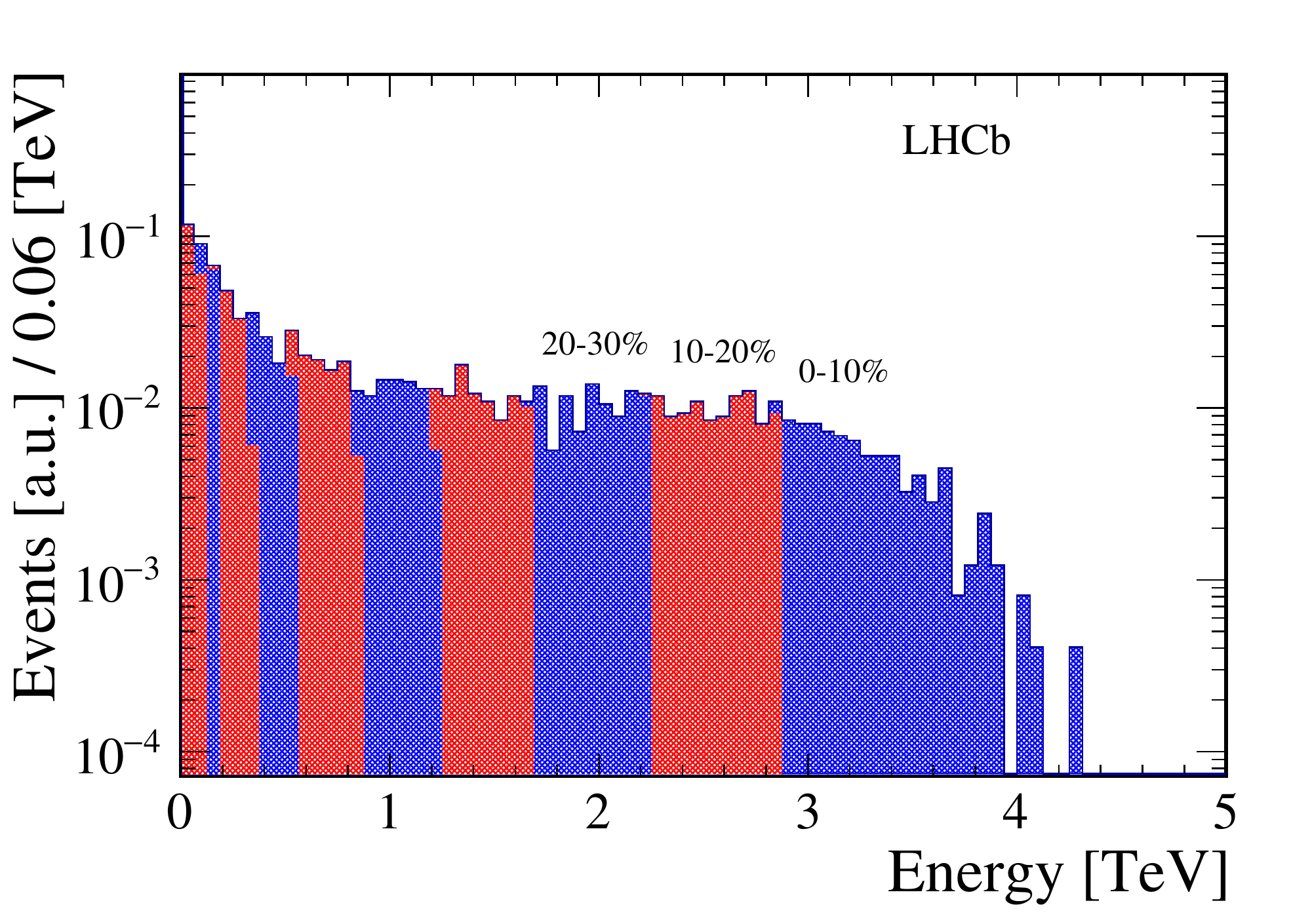}
      \includegraphics[width=0.48\textwidth]{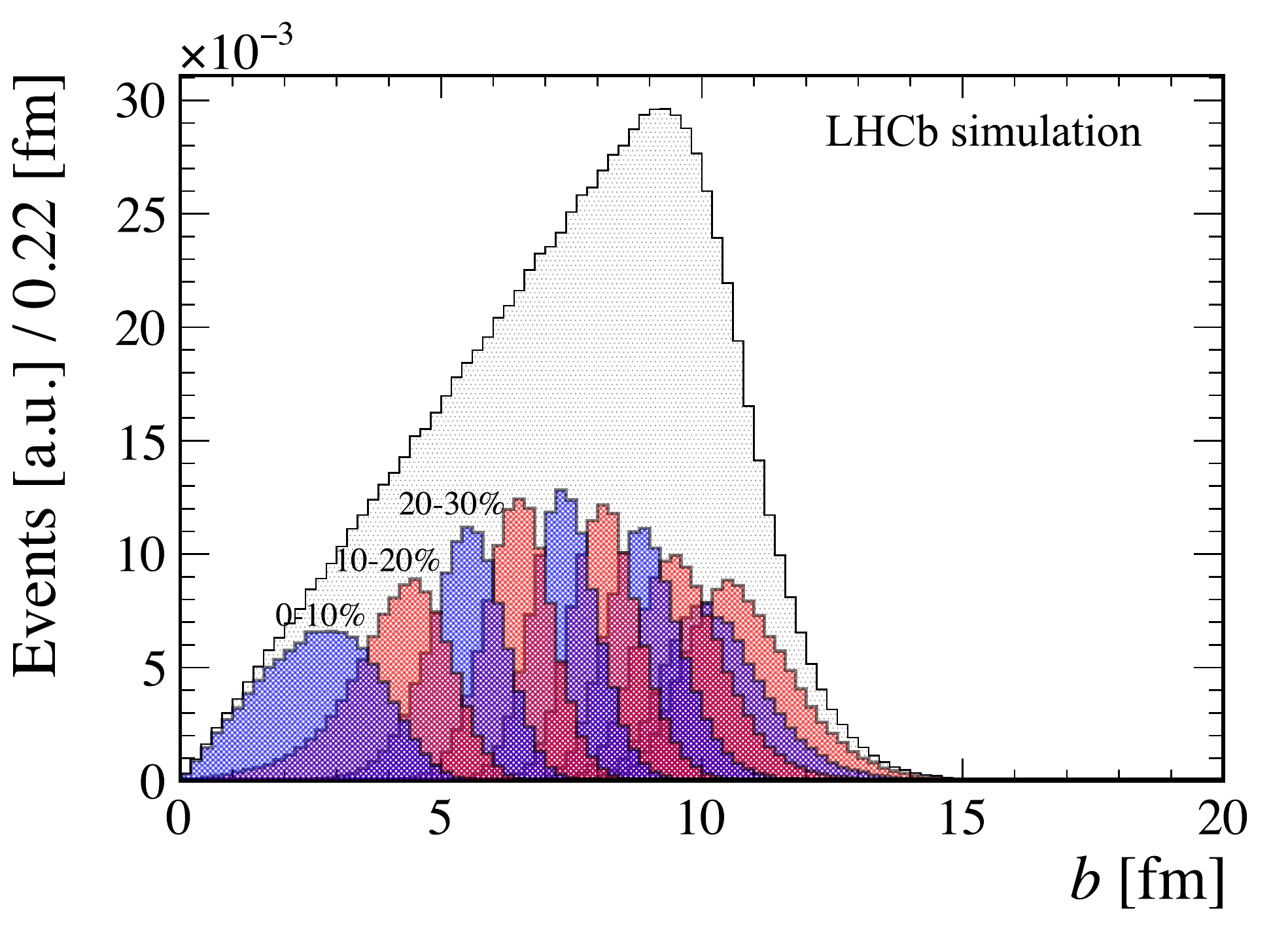}
      \includegraphics[width=0.48\textwidth]{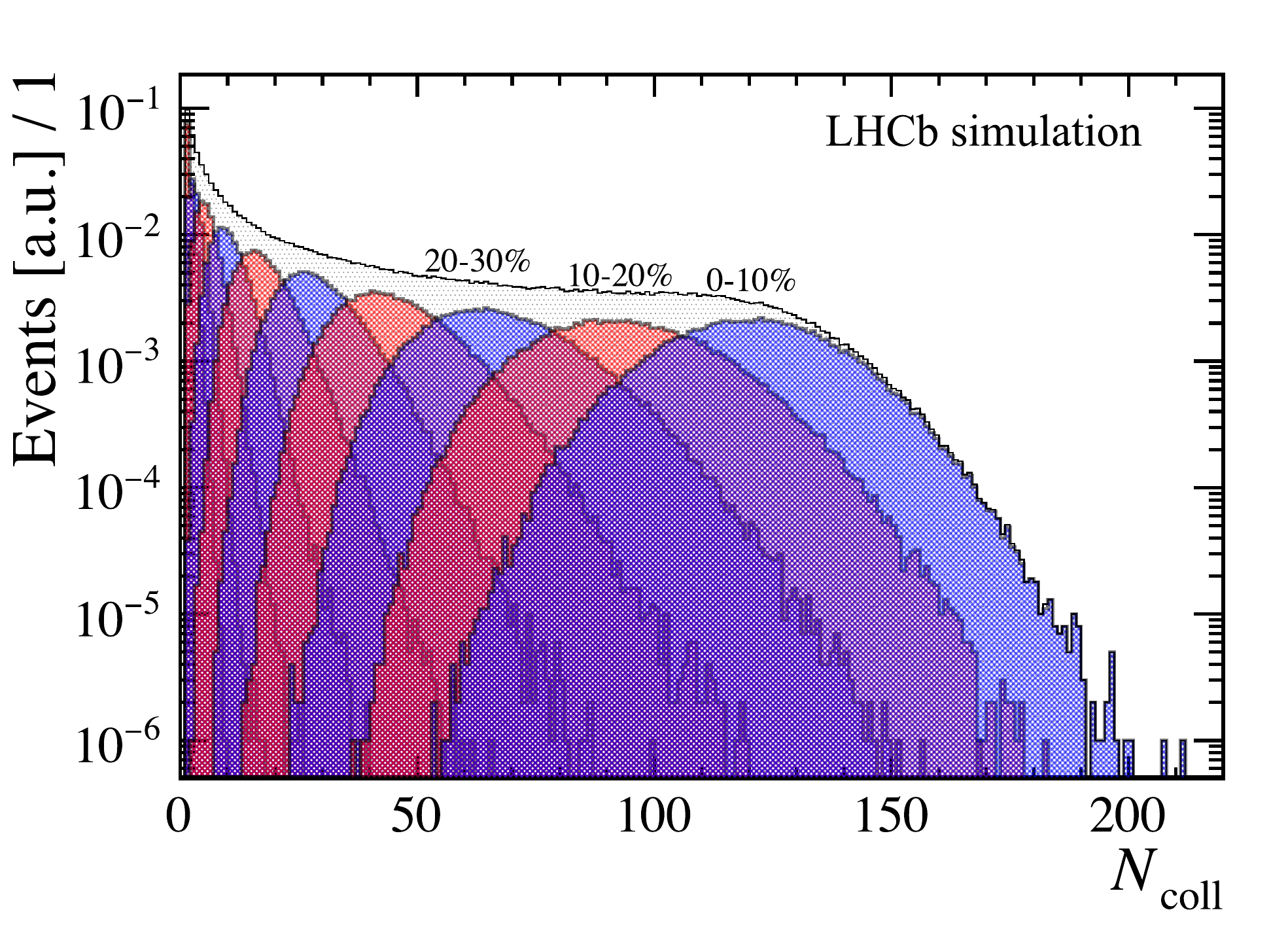}
      \includegraphics[width=0.48\textwidth]{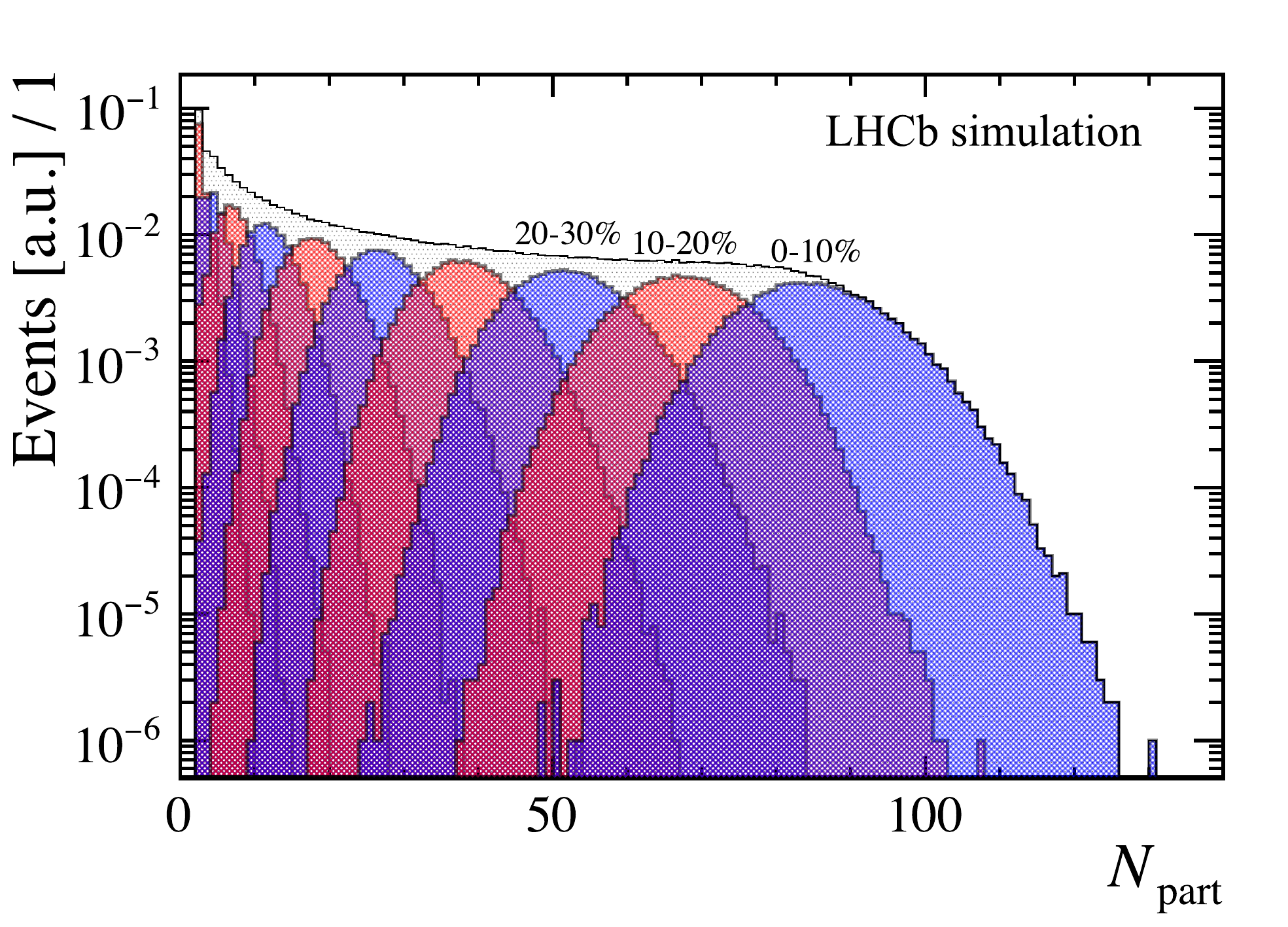}
    \caption{(top left) Classification of events from PbNe data according to the defined centrality classes, distribution of the (top right) impact parameter, (bottom left) \ncoll  and (bottom right) \npart values for the corresponding centrality classes.}
    \label{fig:cent_classes_pbne}
  \end{figure}

The centrality classification of the MB dataset of PbPb collisions in percentile intervals of 10\%, with the cuts in energy obtained as in the previously described procedure, is shown in Fig.~\ref{fig:cent_classes} as well as the $b$, \npart and \ncoll distributions for each class obtained from the Glauber MC model. For each class a mean number is estimated for each of the quantities of interest, together with their corresponding standard deviations. The same distributions for the PbNe case can be seen in Fig.~\ref{fig:cent_classes_pbne}.

In this way, one can define as many classes as desired and of arbitrary width in percentiles. The values of the geometric quantities for each class, as well as the corresponding energy requirements, can be seen in Table~\ref{tab:cent_10} for PbPb and in Table~\ref{tab:cent_10_pbne} for PbNe. Ten classes are shown for each case.

\begin{table}[b]
\caption{Geometric quantities (\npart, \ncoll and $b$) of PbPb collisions for centrality classes defined from a Glauber MC model fitted to the data. The classes correspond to sharp cuts in the energy deposited in the \ecal. Here $\sigma$ stands for the standard deviation of the corresponding distributions.}
\begin{center}
\begin{tabular}{r@{\:$-$\:}l|r@{\:$-$\:}lrrrrrc}

    \multicolumn{2}{c|}{Centrality \%} & \multicolumn{2}{c}{E [\gev]} & \multicolumn{1}{c}{\npart}  & $\sigma_{\npart}$ & \multicolumn{1}{c}{\ncoll}  & $\sigma_{\ncoll}$ &  \multicolumn{1}{c}{$b$}  & $\sigma_{b}$       \\ 
    \hline
     $100$ & $90$ & $0$ & $310$ & $2.9$ & $1.2$ & $1.8$ & $1.2$ & $15.4$ & $1.0$ \\
 $90$ & $80$ & $310$ & $800$ & $7.0$ & $2.9$ & $5.8$ & $3.1$ & $14.6$ & $0.9$ \\
 $80$ & $70$ & $800$ & $1750$ & $15.9$ & $4.8$ & $16.4$ & $7.0$ & $13.6$ & $0.7$ \\
 $70$ & $60$ & $1750$ & $3360$ & $31.3$ & $7.1$ & $41.3$ & $14.7$ & $12.6$ & $0.6$ \\
 $60$ & $50$ & $3360$ & $5900$ & $54.7$ & $10.0$ & $92.6$ & $27.7$ & $11.6$ & $0.5$ \\
 $50$ & $40$ & $5900$ & $9630$ & $87.5$ & $13.3$ & $187.5$ & $46.7$ & $10.5$ & $0.5$ \\
 $40$ & $30$ & $9630$ & $14860$ & $131.2$ & $16.9$ & $345.5$ & $71.6$ & $9.2$ & $0.5$ \\
 $30$ & $20$ & $14860$ & $22150$ & $188.0$ & $21.5$ & $593.9$ & $105.2$ & $7.8$ & $0.6$ \\
 $20$ & $10$ & $22150$ & $32280$ & $261.8$ & $27.1$ & $972.5$ & $151.9$ & $6.0$ & $0.7$ \\
 $10$ & $0$ & $32280$ & $\infty$ & $357.2$ & $32.2$ & $1570.3$ & $236.8$ & $3.3$ & $1.2$
  \end{tabular}\end{center}
\label{tab:cent_10}
\end{table}

\begin{table}[b]
\caption{Geometric quantities (\npart, \ncoll and $b$) of PbNe collisions for centrality classes defined from a MC Glauber model fitted to the data. The classes correspond to sharp cuts in the energy deposited in the \ecal. Here $\sigma$ stands for the standard deviation of the corresponding distributions.}
\begin{center}
\begin{tabular}{r@{\:$-$\:}l|r@{\:$-$\:}lrrrrrc}

    \multicolumn{2}{c|}{Centrality \%} & \multicolumn{2}{c}{E [\gev]} & \multicolumn{1}{c}{\npart}  & $\sigma_{\npart}$ & \multicolumn{1}{c}{\ncoll}  & $\sigma_{\ncoll}$ &  \multicolumn{1}{c}{$b$}  & $\sigma_{b}$       \\ 
    \hline
     $100$ & $90$ & $0$ & $94$ & $2.5$ & $0.8$ & $1.4$ & $0.7$ & $10.9$ & $1.1$ \\
 $90$ & $80$ & $94$ & $184$ & $3.9$ & $1.6$ & $2.7$ & $1.5$ & $10.4$ & $1.0$ \\
 $80$ & $70$ & $184$ & $324$ & $6.8$ & $2.4$ & $5.2$ & $2.4$ & $9.7$ & $0.9$ \\
 $70$ & $60$ & $324$ & $533$ & $11.3$ & $3.2$ & $9.7$ & $3.8$ & $9.0$ & $0.8$ \\
 $60$ & $50$ & $532$ & $828$ & $17.9$ & $4.2$ & $17.3$ & $5.9$ & $8.2$ & $0.7$ \\
 $50$ & $40$ & $828$ & $1213$ & $26.7$ & $5.2$ & $29.0$ & $8.7$ & $7.4$ & $0.6$ \\
 $40$ & $30$ & $1213$ & $1690$ & $38.0$ & $6.3$ & $45.6$ & $12.3$ & $6.5$ & $0.7$ \\
 $30$ & $20$ & $1690$ & $2250$ & $51.7$ & $7.5$ & $67.8$ & $16.1$ & $5.4$ & $0.8$ \\
 $20$ & $10$ & $2250$ & $2879$ & $67.3$ & $8.3$ & $94.1$ & $18.9$ & $4.1$ & $1.0$ \\
 $10$ & $0$ & $2879$ & $\infty$ & $84.8$ & $9.5$ & $120.4$ & $18.6$ & $2.7$ & $1.1$
  \end{tabular}\end{center}
\label{tab:cent_10_pbne}
\end{table}

The PbNe results, when compared to the PbPb case, exhibit a larger uncertainty in the values of the geometrical quantities. This limited precision is an effect of the system size which is much smaller in the case of PbNe, and not a result of the use of the \ecal to estimate centrality. If the centrality classes were defined purely from the Glauber model by applying sharp cuts in the $b$ distribution, the same large overlaps in the \ncoll and \npart distributions would be found, without ever including the \ecal in the procedure.

One important caveat is that at low energy the dominating events are of electromagnetic nature or from UPC. Because of this, it is important to exclude in the analyses the energy region where there is a sizeable contamination from these events. If no further selection has been applied to reject UPC events, its contamination in PbPb events will be below 5\% at energies higher than 585\gev, that is at centralities lower than 84\% (more central than 84\%), and for PbNe at energies higher than $98.9\gev$, that is at centralities lower than 89\%. 

To determine this threshold, the data are compared to the fitted Glauber MC as in the right plot
of Fig.~\ref{fig:fit_to_data}. 
The point from which the two distributions match is found by computing a centred mean of the data over MC ratio around each bin. When this ratio is below a chosen tolerance of 1.05 (meaning 5\% contamination of UPC events) for three consecutive bins, the centre of the bin of lower energy is chosen as the energy threshold.\footnote{For PbPb, looking at a given bin $n$, the ratio of data over MC is computed for bins $n-1$, $n$ and $n+1$, and averaged. For PbNe, the ratio of data over MC is computed for the five precedent bins, for bin $n$, and for the five subsequent bins, and averaged.}
If UPC events were identified and rejected, then this limit, of 84\% for PbPb or 89\% for PbNe, could be increased to include more peripheral events.

The results in the PbPb case are in very good agreement with the results obtained by the ALICE~\cite{PhysRevLett.116.222302, ALICE-PUBLIC-2015-008}, \atlas~\cite{ATLAS2019108} and \cms~\cite{PhysRevLett.127.102002} collaborations at the same centre-of-mass energy. The PbNe results correspond to the first centrality measurements in fixed-target collisions at the LHC.

%% file: uncertainties.tex
\section{Systematic uncertainties}
\label{sec.uncertainties}

In the given centrality classes with fixed deposited energy boundaries, the uncertainties on the geometric quantities, like the mean values of \npart, \ncoll and $b$, are assessed. The previously found energy selections are kept, but the systematic uncertainties are quantified on the geometric properties. In what follows, the systematic uncertainties are reported in tables of ten classes of ten percentiles each. 

\subsection{Bin-width dependence}

To find the boundary values of the ECAL energy bins, an integration procedure is performed on the histogram of the simulated energy deposition per event. The larger the bins, the less precise the percentile of events in the energy bins can be determined.

A binning scheme of 6000 bins is used, which has an average miss percentage of $0.04\%$ for the PbPb case and $0.02\%$ for PbNe. Since each percentile corresponds to a 1\% interval, an average miss of 0.04\% means that on average 4\% of the events of one percentile ``migrate'' to the class immediately below in energy. The effect on the geometric uncertainties is estimated and the results are stored for every percentile and \npart, \ncoll and $b$.
  
\subsection{Hadronic cross-section uncertainty}

One of the main ingredients for the Glauber MC model is the nucleon-nucleon cross-section. For PbPb at a centre-of-mass energy of $\sqsnn = 5\tev$, it corresponds to \mbox{$\siginel =67.6 \pm 0.6 \mbarn$}, where the uncertainty comes from the data driven parametrisation described in Ref.~\cite{PhysRevC.97.054910}. For PbNe at a centre-of-mass energy of $\sqsnn = 69\gev$, the nucleon-nucleon cross-section corresponds to \mbox{$\siginel =35.4 \pm 0.9 \mbarn$}. To quantify the effect this uncertainty has in the geometric properties, 
the simulated energy distribution is
generated from a Glauber MC simulation made with  $\siginel+1\sigma$ and with $\siginel-1\sigma$ for each collision system. Then the centrality classes are defined with these new distributions and the effect on the mean values for \npart, \ncoll and $b$ is taken as the associated systematic uncertainty.

\subsection{Fit uncertainty}

One of the most important steps in the process of determining centrality is the choice of the parameters $f$ and $\mu$. From Sect.~\ref{seq:fittodata}, for PbPb there are three best fits that are found, and four for PbNe. These best fits are used to compute the systematic uncertainties due to the choice of a given set of $(f,\mu)$ for PbPb and then for PbNe. 

In order to compute the uncertainty the Glauber MC energy distribution is generated with the different sets of values. For both PbPb and PbNe separately,  the centrality classes are defined for each set of best fits and finally the resulting mean values of the geometric quantities (\npart, \ncoll and $b$) are compared between the best fits that were not kept and the one that was kept as the definitive best fit.

The precision at which the mean energy deposited per particle is determined in the \ecal does not affect the final result, since any variation of this value would be compensated by the value of $\mu$ found. Thus the uncertainty on this energy is absorbed into the uncertainty due the choice of a given set of parameters $(f,\mu)$.

\subsection{NBD uncertainty}

The NBD sampling introduces statistical fluctuations that affect directly the observed \chisq value when comparing the  Glauber MC simulation and the data. This effect is noticeable even when looking at the same point of the $(f,\mu)$ parameter space. To estimate how these fluctuations affect the final computed geometric quantities, ten simulated energy distributions are 
simulated with the same best fit parameters and the geometric quantities are computed for all of them. 

For each percentile, the standard deviation for \npart, \ncoll and $b$ is computed  and used as uncertainty. As before, this is done separately for PbPb and for PbNe.

\subsection{Total systematic uncertainties}

These uncertainties are added together in quadrature to obtain the total uncertainty for each centrality class. The result can be seen in Table~\ref{tab:tot_syst_10} for ten classes for the PbPb case, and in Table~\ref{tab:tot_syst_10_pbne} for the PbNe case.

The uncertainties on the geometric quantities in both cases are dominated by the systematic uncertainties, as expected. In the PbPb case, the dominant one is the uncertainty due to the binning effect, while in the PbNe case, the dominant one is the uncertainty due to the binning effect in more peripheral collisions (centrality higher than 50\%) and the uncertainty due to the hadronic cross-section uncertainty for more central events (centrality lower than 50\%). 

\begin{table}[]
\caption{Total uncertainties for the geometric quantities (\npart, \ncoll and $b$) of PbPb collisions for centrality classes defined from a MC Glauber model fit to the data. The statistical and systematic uncertainties are added in quadrature.}
\begin{center}
\begin{tabular}{r@{\:$-$\:}l|r@{\:$\pm$\:}lr@{\:$\pm$\:}lr@{\:$\pm$\:}l}

    \multicolumn{2}{c|}{Centrality \%} & \npart  & $\sigma$ & \ncoll  & $\sigma$ &  $b$  & $\sigma$       \\ 
    \hline
$100$ & $90$ & $2.91$ & $0.54$ & $1.83$ & $\phantom{0}0.34$ & $15.41$ & $2.96$ \\ 
$90$ & $80$ & $7.03$ & $0.78$ & $5.77$ & $\phantom{0}0.64$ & $14.56$ & $1.80$ \\ 
$80$ & $70$ & $15.92$ & $0.64$ & $16.44$ & $\phantom{0}0.69$ & $13.59$ & $0.52$ \\ 
$70$ & $60$ & $31.26$ & $0.67$ & $41.28$ & $\phantom{0}0.93$ & $12.61$ & $0.28$ \\ 
$60$ & $50$ & $54.65$ & $1.13$ & $92.59$ & $\phantom{0}2.01$ & $11.59$ & $0.24$ \\ 
$50$ & $40$ & $87.54$ & $1.01$ & $187.54$ & $\phantom{0}2.43$ & $10.47$ & $0.14$ \\ 
$40$ & $30$ & $131.24$ & $1.15$ & $345.53$ & $\phantom{0}3.89$ & $9.23$ & $0.08$ \\ 
$30$ & $20$ & $188.02$ & $1.49$ & $593.92$ & $\phantom{0}6.62$ & $7.80$ & $0.06$ \\ 
$20$ & $10$ & $261.84$ & $1.83$ & $972.50$ & $10.37$ & $6.02$ & $0.04$ \\ 
$10$ & $0$ & $357.16$ & $1.70$ & $1570.26$ & $15.56$ & $3.31$ & $0.01$

  \end{tabular}\end{center}
\label{tab:tot_syst_10}
\end{table}

\begin{table}[]
\caption{Total uncertainties for the geometric quantities (\npart, \ncoll and $b$) of PbNe collisions for centrality classes defined from a MC Glauber model fit to the data. The statistical and systematic uncertainties are added in quadrature.}
\begin{center}
\begin{tabular}{r@{\:$-$\:}l|r@{\:$\pm$\:}lr@{\:$\pm$\:}lr@{\:$\pm$\:}l}

    \multicolumn{2}{c|}{Centrality \%} & \npart  & $\sigma$ & \ncoll  & $\sigma$ &  $b$  & $\sigma$       \\ 
    \hline
$100$ & $90$ & $2.45$ & $0.07$ & $1.41$ & $0.04$ & $10.85$ & $0.33$ \\ 
$90$ & $80$ & $3.93$ & $0.15$ & $2.67$ & $0.11$ & $10.37$ & $0.41$ \\ 
$80$ & $70$ & $6.80$ & $0.25$ & $5.21$ & $0.20$ & $9.69$ & $0.34$ \\ 
$70$ & $60$ & $11.34$ & $0.28$ & $9.73$ & $0.27$ & $8.95$ & $0.22$ \\ 
$60$ & $50$ & $17.88$ & $0.25$ & $17.25$ & $0.36$ & $8.19$ & $0.09$ \\ 
$50$ & $40$ & $26.72$ & $0.30$ & $28.95$ & $0.62$ & $7.38$ & $0.06$ \\ 
$40$ & $30$ & $37.99$ & $0.55$ & $45.62$ & $1.11$ & $6.48$ & $0.08$ \\ 
$30$ & $20$ & $51.72$ & $0.56$ & $67.78$ & $1.60$ & $5.44$ & $0.03$ \\ 
$20$ & $10$ & $67.30$ & $0.78$ & $94.10$ & $2.32$ & $4.14$ & $0.03$ \\ 
$10$ & $0$ & $84.84$ & $0.98$ & $120.43$ & $3.04$ & $2.67$ & $0.03$

  \end{tabular}\end{center}
\label{tab:tot_syst_10_pbne}
\end{table}

%% file: conclusions.tex
\section{Conclusions}

A procedure to determine the centrality in PbPb collisions at $\sqsnn =5\tev$ and in PbNe collisions at $\sqsnn =69\gev$ with the \lhcb detector is implemented. The distributions of measured energy deposits in the ECAL are fitted to obtain the parameters of the Glauber model for the simulation. After the fit is performed, the simulated distribution is divided in percentiles, which are delimited by sharp energy boundaries obtained by integrating the distribution. These energy selections allows to classify the data into the same percentiles and subsequently the geometric quantities from the Glauber MC model can be mapped to the real data. The obtained centrality classification is limited to the 84\% (89\%) most central PbPb (PbNe) events, avoiding the region with large contamination from ultra-peripheral collisions. The correspondence between the results obtained for the PbPb collisions is in good agreement with the results from the \alice, \atlas and \cms experiments, and the centrality measurements for the PbNe collisions presented here are the first performed in fixed-target collisions at the LHC.

%% file: acknowledgements.tex
\section*{Acknowledgements}
%
%
\noindent We express our gratitude to our colleagues in the CERN
accelerator departments for the excellent performance of the LHC. We
thank the technical and administrative staff at the LHCb
institutes.
We acknowledge support from CERN and from the national agencies:
CAPES, CNPq, FAPERJ and FINEP (Brazil); 
MOST and NSFC (China); 
CNRS/IN2P3 (France); 
BMBF, DFG and MPG (Germany); 
INFN (Italy); 
NWO (Netherlands); 
MNiSW and NCN (Poland); 
MEN/IFA (Romania); 
MSHE (Russia); 
MICINN (Spain); 
SNSF and SER (Switzerland); 
NASU (Ukraine); 
STFC (United Kingdom); 
DOE NP and NSF (USA).
We acknowledge the computing resources that are provided by CERN, IN2P3
(France), KIT and DESY (Germany), INFN (Italy), SURF (Netherlands),
PIC (Spain), GridPP (United Kingdom), RRCKI and Yandex
LLC (Russia), CSCS (Switzerland), IFIN-HH (Romania), CBPF (Brazil),
PL-GRID (Poland) and NERSC (USA).
We are indebted to the communities behind the multiple open-source
software packages on which we depend.
Individual groups or members have received support from
ARC and ARDC (Australia);
AvH Foundation (Germany);
EPLANET, Marie Sk\l{}odowska-Curie Actions and ERC (European Union);
A*MIDEX, ANR, IPhU and Labex P2IO, and R\'{e}gion Auvergne-Rh\^{o}ne-Alpes (France);
Key Research Program of Frontier Sciences of CAS, CAS PIFI, CAS CCEPP, 
Fundamental Research Funds for the Central Universities, 
and Sci. \& Tech. Program of Guangzhou (China);
RFBR, RSF and Yandex LLC (Russia);
GVA, XuntaGal and GENCAT (Spain);
the Leverhulme Trust, the Royal Society
 and UKRI (United Kingdom).

%% file: Authorship_LHCb-DP-2021-002.tex
\centerline
{\large\bf LHCb collaboration}
\begin
{flushleft}
\small
R.~Aaij$^{32}$,
C.~Abell{\'a}n~Beteta$^{50}$,
T.~Ackernley$^{60}$,
B.~Adeva$^{46}$,
M.~Adinolfi$^{54}$,
H.~Afsharnia$^{9}$,
C.A.~Aidala$^{85}$,
S.~Aiola$^{25}$,
Z.~Ajaltouni$^{9}$,
S.~Akar$^{65}$
% ...

{\footnotesize \it

% ...
$^{9}$Universit{\'e} Clermont Auvergne, CNRS/IN2P3, LPC, Clermont-Ferrand, France\\
% ...
$^{25}$INFN Sezione di Milano, Milano, Italy\\
% ...
$^{32}$Nikhef National Institute for Subatomic Physics, Amsterdam, Netherlands\\
% ...
$^{46}$Instituto Galego de F{\'\i}sica de Altas Enerx{\'\i}as (IGFAE), Universidade de Santiago de Compostela, Santiago de Compostela, Spain\\
% ...
$^{50}$Physik-Institut, Universit{\"a}t Z{\"u}rich, Z{\"u}rich, Switzerland\\
% ...
$^{54}$H.H. Wills Physics Laboratory, University of Bristol, Bristol, United Kingdom\\
% ...
$^{60}$Oliver Lodge Laboratory, University of Liverpool, Liverpool, United Kingdom\\
% ...
$^{65}$University of Cincinnati, Cincinnati, OH, United States\\
% ...
$^{68}$Syracuse University, Syracuse, NY, United States\\
% ...
$^{85}$University of Michigan, Ann Arbor, United States, associated to $^{68}$\\
\bigskip
% ...
\medskip
}
\end{flushleft}